\documentclass[english,structabstract]{aa}
\usepackage{mathptmx}
\usepackage[T1]{fontenc}
\usepackage[utf8]{inputenc}
\usepackage{babel}
\usepackage{url}
\usepackage{graphicx}
\usepackage[unicode=true]
 {hyperref}
\usepackage{breakurl}

\makeatletter

\newcommand{\lyxdot}{.}

\bibpunct{(}{)}{;}{a}{}{,} % to follow the A&A style
\usepackage{hyperref}
\usepackage{txfonts}
\titlerunning{The Stagger-grid -- IV. Limb darkening coefficients}
\authorrunning{Z. Magic et al.}
\newcommand{\teff}{T_{\mathrm{eff}}}
\newcommand{\logg}{\log g}
\newcommand{\feh}{\left[\mathrm{Fe}/\mathrm{H}\right]}
\newcommand{\nab}{\vec{\nabla}}

\newcommand{\hav}{\left\langle \mathrm{3D}\right\rangle}

\@ifundefined{showcaptionsetup}{}{%
 \PassOptionsToPackage{caption=false}{subfig}}
\usepackage{subfig}
\makeatother

\begin{document}

\title{The \textsc{Stagger}-grid: A grid of 3D stellar atmosphere models}

\subtitle{IV. Limb darkening coefficients%
\thanks{Appendix is available in electronic form at \protect\href{http://www.aanda.org}{http://www.aanda.org}%
}%
\thanks{Full Table A.1 is available at the CDS via anonymous ftp to \protect\href{http://cdsarc.u-strasbg.fr}{cdsarc.u-strasbg.fr}
(\protect\href{http://130.79.128.5}{130.79.128.5}) or via \protect\href{http://cdsarc.u-strasbg.fr/viz-bin/qcat?J/A+A/???/A??}{http://cdsarc.u-strasbg.fr/viz-bin/qcat?J/A+A/???/A??},
as well as at \protect\href{http://www.stagger-stars.net}{www.stagger-stars.net}%
}}

\author{Z. Magic\inst{1,2}, A. Chiavassa,\inst{3} R. Collet\inst{2} \and 
M. Asplund\inst{2}}

\institute{Max-Planck-Institut f{\"u}r Astrophysik, Karl-Schwarzschild-Str.
1, 85741 Garching, Germany \\
\email{magic@mpa-garching.mpg.de} \and  Research School of Astronomy
\& Astrophysics, Cotter Road, Weston ACT 2611, Australia\and  Laboratoire
Lagrange, UMR 7293, CNRS, Observatoire de la C{\^o}te d’Azur, Universit{\'e}
de Nice Sophia-Antipolis, Nice, France}

\offprints{magic@mpa-garching.mpg.de}

\date{Received ...; Accepted...}

\abstract{}{ We compute the emergent stellar spectra from the UV to far infrared
for different viewing angles using realistic 3D model atmospheres
for a large range in stellar parameters to predict the stellar limb
darkening. }{We have computed full 3D LTE synthetic spectra based
on 3D radiative hydrodynamic atmosphere models from the \textsc{Stagger}-grid
in the ranges: $T_{\mathrm{eff}}$ from $4000$ to $7000\,\mathrm{K}$,
$\log g$ from $1.5$ to $5.0$, and $\left[\mathrm{Fe}/\mathrm{H}\right]$,
from $-4.0$ to $+0.5$. From the resulting intensities at different
wavelength, we derived coefficients for the standard limb darkening
laws considering a number of often-used photometric filters. Furthermore,
we calculated theoretical transit light curves, in order to quantify
the differences between predictions by the widely used 1D model atmosphere
and our 3D models. }{The 3D models are often found to predict steeper
darkening towards the limb compared to the 1D models, mainly due to
the temperature stratifications and temperature gradients being different
in the 3D models compared to those predicted with 1D models based
on the mixing length theory description of convective energy transport.
The resulting differences in the transit light curves are rather small;
however, these can significant for high-precision observations of
extrasolar transits, and are able to lower the residuals from the
fits with 1D limb darkening profiles. }{We advocate the use of the
new limb darkening coefficients provided for the standard four-parameter
non-linear power law, which can fit the limb darkening more accurately
than other choices.}


\maketitle

\section{Introduction\label{sec:Introduction}}

The emergent intensity across the surface of late-type stars diminishes
gradually from the center of the stellar disk towards the edge (limb),
since the optical depth depends on the angle of view. Rays crossing
the stellar photosphere near the limb reach optical depth unity in
layers at higher altitude and at typically lower densities and temperatures
than rays crossing the stellar disk near the center. The intensity
is very sensitive to the temperature, therefore, one observes darker
brightness from the limb, which emerges from higher and cooler regions
of the stellar atmosphere. This effect is known as limb darkening
\citep{Gray2005oasp.book.....G}. An accurate knowledge of the surface
brightness distribution is essential for the analysis of light curves
from stars with transiting objects in the line of sight, such as exoplanets
and eclipsing stellar companions in binary systems. Furthermore, the
precise determination of stellar angular diameters with stellar interferometry
relies also on the theoretical limb darkening predictions \citep{Davis2000MNRAS.318..387D}.
The variation in surface intensity with angular distance from the
stellar disk center is usually expressed in the form of limb darkening
laws \citep{Claret:2000p12465}. Multiple functional basis have been
used in the past, from simple linear to higher order non-linear laws,
in order to fit the surface brightness variations predicted by theoretical
model atmospheres leading to so-called limb darkening coefficients
(LDC). For example, the individual shape of a light curve for transiting
exoplanets is important, because it contains information about the
structure of the external layers of the occulted stellar object \citep[e.g.,][]{Southworth:2008p23114}.
The observed light curves are interpreted by comparisons with theoretical
transit light curves that are based on limb darkening predictions
arising from model atmospheres. More accurate theoretical atmosphere
models will reduce the uncertainties in the comparison, and thereby
improving the quality of the analysis in favor of other transit-parameters
like planet-to-star ratio or the inclination of the orbit. Also, the
goodness of the transmission-spectroscopy of exoplanet atmospheres
relies on the underlying theoretical atmospheres of the host stars
\citep[e.g.,][]{Seager:2000p14757}.

The first estimates of the intensity variation over the disk were
performed with a simple linear law \citep{Milne1921MNRAS..81..361M}.
However, with theoretical 1D model atmospheres it was shown that a
linear law is insufficient to describe the limb darkening of a real
star adequately \citep[e.g.,][]{vanHamme:1993p22403}. Then, various
alternatives with a two-parameter law was introduced starting from
a quadratic, over square root to a logarithmic, and finally an exponential
law \citep[e.g. see][]{DiazCordoves:1995p22374,Claret:1995p22394}.
These restricted functional bases are only marginally accurate for
a certain range in effective temperatures, therefore, \citet{Claret:2000p12465}
introduced a new non-linear power law with four coefficients, which
is powerful enough to fit the LDC for a broad range in stellar parameters,
while conserving the flux to a high accuracy. Later on, limb darkening
variations were fitted and provided for the community derived from
extensive grids with the latest model atmospheres (e.g, \textsc{MARCS,
ATLAS} and \textsc{PHOENIX}) for several broad band filters, e.g.
the SDSS \citep{Claret:2004p12494}, Kepler and CoRoT \citep{Sing:2010p21961}.
An extensive comparison of the various limb darkening laws has been
performed by \citet{Southworth:2008p23114}. All of these developments
revealed that a well-considered choice of an appropriate functional
basis is mandatory for a precise description of the intensity variations.

The next step in improving the systematic errors prevailing in the
predicted limb darkening laws was yielded in the underlying model
atmospheres, since the limb darkening is mainly determined by the
temperature gradient \citep[see][]{Knutson:2007p22139,Hayek:2012p21944}.
Therefore, flaws in the theoretical atmospheric temperature stratification
will directly propagate into the predicted limb darkening. The hydrostatic
1D models make use of several simplifications, the most prominent
one being the use of the mixing length theory to account for convective
energy transport \citep{BohmVitense:1958p4822}. Cool late-type stars
feature a convective envelope, thereby convective motions are present
in the thin photospheric transition region due to overshooting of
convective flows. These stars exhibit a typical granulation pattern
in its emergent intensity due to inhomogeneities arising from the
asymmetric up- and downflowing stellar plasma. Therefore, only 3D
atmosphere models are able to predict these properties accurately.
With the advent of 3D atmosphere modeling \citep{Nordlund:1982p6697},
which solves from first-principle the hydrodynamic equations coupled
with a realistic radiative transfer, the deficiencies of the 1D models
were revealed and quantified \citep[e.g.,][ and references therein]{Nordlund:2009p4109}.
Comparisons of the 3D models with the Sun showed that these models
can predict accurately the intensity distribution \citep{Pereira:2013arXiv1304},
while 1D models overestimate the limb darkening of our resolved host
star. \citet{Bigot:2006p23513} studied the limb darkening of $\alpha$
Centauri B by comparing its interferometrically observed visibility
curves with theoretical predictions. The latter is sensitive to the
limb darkening, and they found an significant improvement with the
predictions from 3D models. Furthermore, \citet{Hayek:2012p21944}
showed on the basis of the extremely accurately measured light curves
of the transiting exoplanet HD 209458 that the intrinsic residuals
of the 1D models can be resolved with the more realistic 3D model
atmospheres. The largest differences were found close to the limb,
hence during the ingress and egress of the transition. With 1D model
predictions, the well-studied close-orbit Jupiter-like transit planet
HD 209458 exhibited priorly systematic residuals due to the simplified
treatment of convection leading to insufficient temperature stratifications
\citep[see][]{Knutson:2007p22139}. For another well-studied star,
Procyon, and four K giants \citet{Chiavassa:2010p6257,Chiavassa:2012p22493}
could also find the limb darkening and stellar diameter predictions
to be coherent with independent asteroseismic observations \citep[see also][]{AllendePrieto:2002p22280,Aufdenberg2005ApJ...633..424A}. 

After the first detection of a Jupiter-like extra solar planet through
radial velocity detection \citep{Mayor:1995p23118}, five years later,
eventually a transiting exoplanet around a solar-like star was also
found \citep{Charbonneau:2000p22468}. These spectacular landmark
discoveries triggered literally a gold-rush in the hunt for new exoplanets.
With advanced satellite missions, like Kepler and CoRoT, nowadays
up $1491$ transiting extra solar planets have been detected \citep{Wright2011PASP..123..412W}.
Both of the mentioned satellite missions operate in the visible spectral
range, therefore, the effects of limb darkening are strong. These
sophisticated observations evoke rightfully a demand in more accurate
theoretical limb darkening predictions. In order to fulfill this call,
we present in this work LDCs derived from realistic full 3D synthetic
spectra based on a comprehensive grid of 3D RHD atmosphere models.

In Sects. \ref{sec:Methods} and \ref{sec:Deriving-limb-darkening},
we explain the methods we utilized to obtain the LDC. Subsequently,
the resulting theoretical limb darkening variations (Sec. \ref{sec:Limb-darkening-laws})
and transit light curves (Sec. \ref{sec:Transit-light-curves}) are
presented and discussed. We compare our results with previous predictions
from 1D ATLAS models in Sec. \ref{sec:Comparison-with-1D}. Finally,
we conclude our findings in Sec. \ref{sec:Conclusions}.

\section{3D atmosphere models and 3D synthetic spectra\label{sec:Methods}}

We have computed a large grid of realistic 3D atmosphere models \citep[see][hereafter Paper I]{Magic:2013}.
We employed the \textsc{Stagger}-code, a state-of-the-art (magneto)hydrodynamic
code that solves the time-dependent equations for conservation of
mass, momentum and energy. In the optically thin regime, the code
solves the radiative transfer for the vertical direction and eight
inclined rays along long characteristics \citep{Nordlund:1982p6697,Stein:1998p3801}.
The original large set of wavelength points for the opacity sampling
data are grouped together into 12 opacity bins in the so-called opacity
binning method \citep{Nordlund:1982p6697,Skartlien:2000p9857} while
solving the radiative transfer. The \textsc{Stagger}-code utilizes
a realistic EOS \citep{Mihalas:1970p21310} and continuum and line
opacities (\citealp{Kurucz:1979p4707} with subsequent updates; \citealp{Gustafsson:2008p3814}).
We perform so-called \textquotedbl{}box-in-the-star\textquotedbl{}
simulations, where only a small representative volume is considered
that accommodates the top of the convection zone and the extended
photosphere. The vertical directions feature open boundaries, while
the horizontal ones are periodic. The numerical resolution of the
geometrical mesh is $240^{3}$. The horizontal mesh is equidistant,
while the vertical depth scale is optimized to resolve the photospheric
transition region with an enhanced resolution, thereby exploiting
the given resolution best possibly. The \textsc{Stagger}-grid covers
a wide range in stellar parameters with effective temperatures from
$4000\,\mathrm{K}$ to $7000\,\mathrm{K}$ in steps of $500\,\mathrm{K}$,
surface gravities from $1.5$ to $5.0$ in steps of $0.5$ and metallicities
from $-4.0$ to $+0.5$ in steps of $1.0$ below $-1.0$, and steps
of $0.5$ above. The range in surface gravity and metallicity for
the \textsc{Stagger}-grid covers the range of the planet hosting stars,
while the effective temperature is slightly smaller. For further details
on the model atmospheres we refer to Paper I. Furthermore, in \citet{magic:2013arXiv1307.3273M}
we explain in detail the horizontal averaging methods for the $\hav$
models. Furthermore, additional results on the \textsc{Stagger}-grid
are given in \citet{Magic2014arXiv1403.1062M,Magic2014arXiv1403.6245M}
and \citet{Magic2014arXiv1405.7628M}.

\begin{figure*}
\hspace{7mm}$\teff=5777\, K,\,\logg=4.44$ \hspace{7mm}$\teff=6500\, K,\,\logg=4.00$\hspace{7mm}$\teff=4500\, K,\,\logg=2.00$\hspace{7mm}$\teff=4500\, K,\,\logg=5.00$

\includegraphics[angle=90,width=45mm]{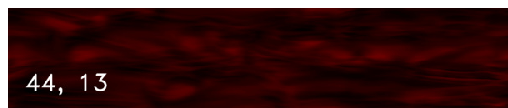}~\includegraphics[angle=90,width=45mm]{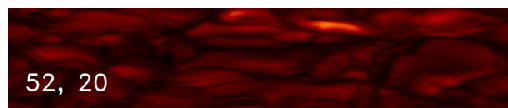}~\includegraphics[angle=90,width=45mm]{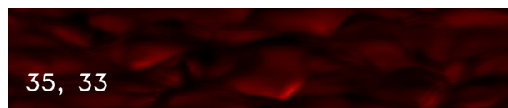}~\includegraphics[angle=90,width=45mm]{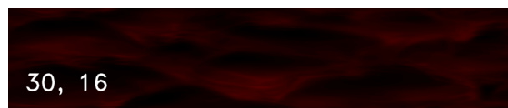}

\vspace{0.001mm}
\includegraphics[angle=90,width=45mm]{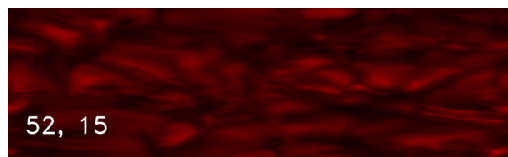}~\includegraphics[angle=90,width=45mm]{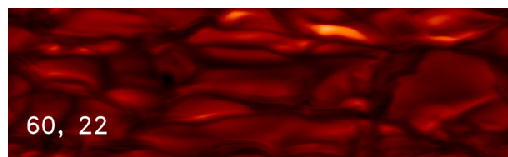}~\includegraphics[angle=90,width=45mm]{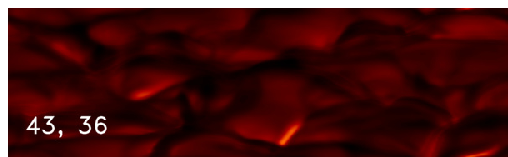}~\includegraphics[angle=90,width=45mm]{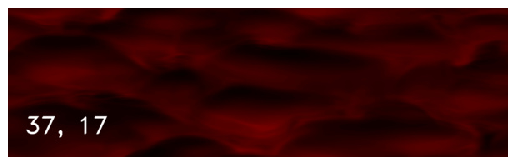}

\vspace{0.001mm}
\includegraphics[angle=90,width=45mm]{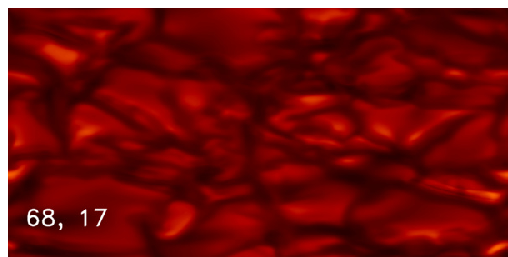}~\includegraphics[angle=90,width=45mm]{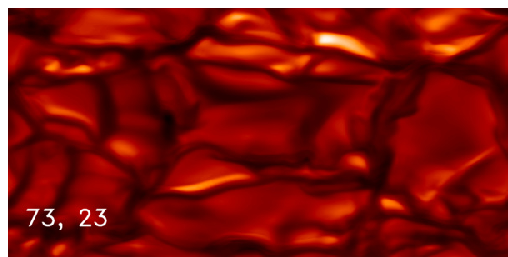}~\includegraphics[angle=90,width=45mm]{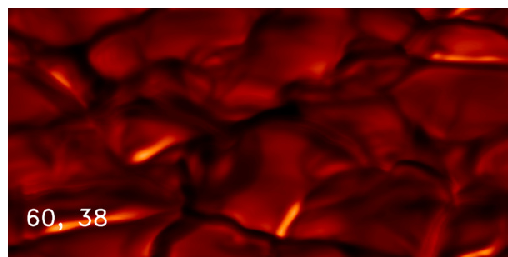}~\includegraphics[angle=90,width=45mm]{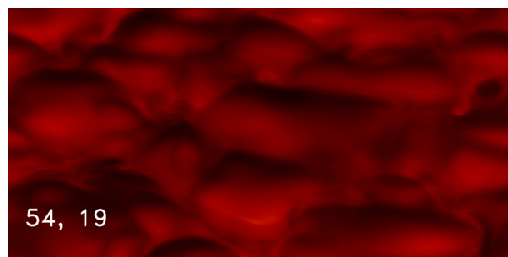}

\vspace{0.001mm}
\includegraphics[angle=90,width=45mm]{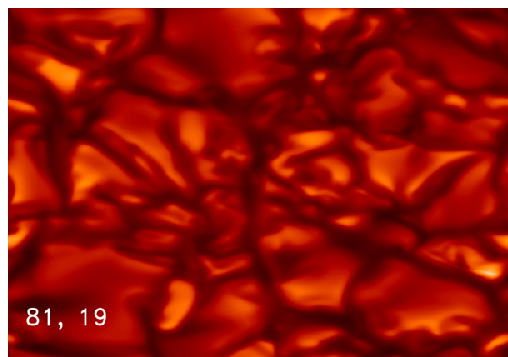}~\includegraphics[angle=90,width=45mm]{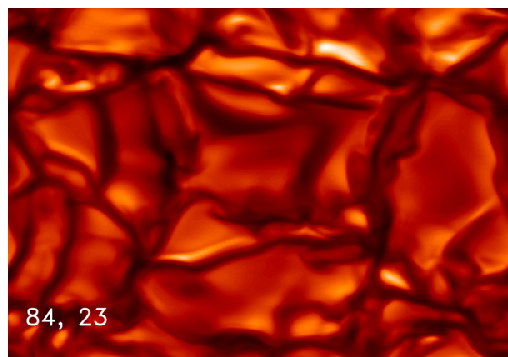}~\includegraphics[angle=90,width=45mm]{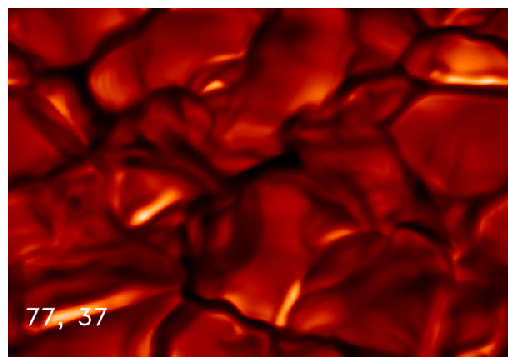}~\includegraphics[angle=90,width=45mm]{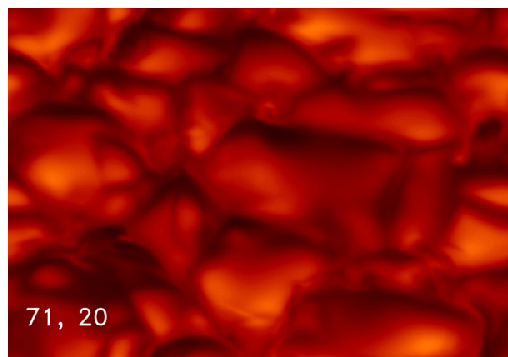}

\vspace{0.001mm}

\includegraphics[angle=90,width=45mm]{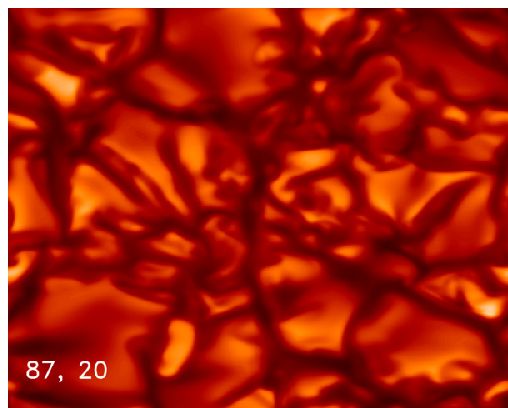}~\includegraphics[angle=90,width=45mm]{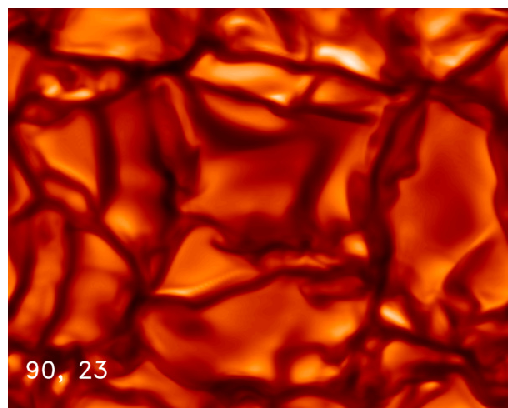}~\includegraphics[angle=90,width=45mm]{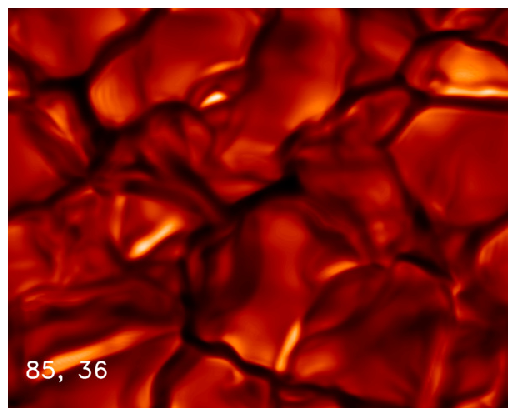}~\includegraphics[angle=90,width=45mm]{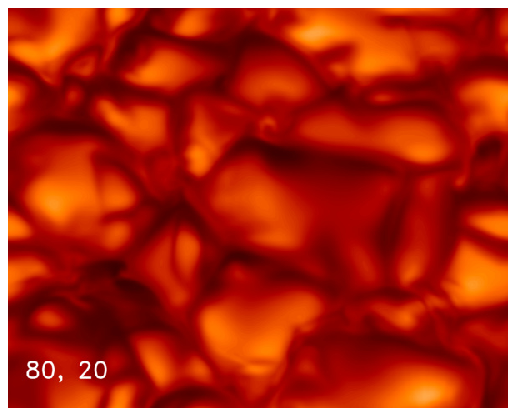}

\vspace{0.001mm}

\includegraphics[angle=90,width=45mm]{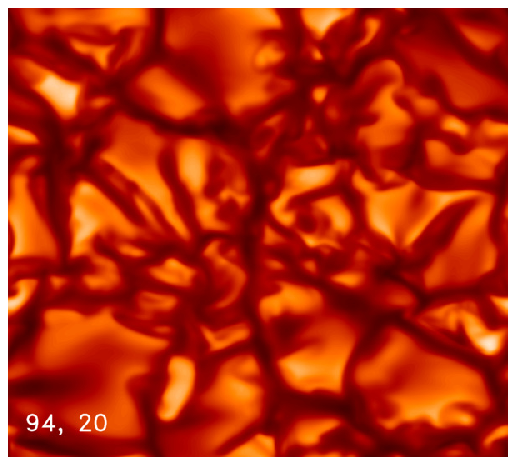}~\includegraphics[angle=90,width=45mm]{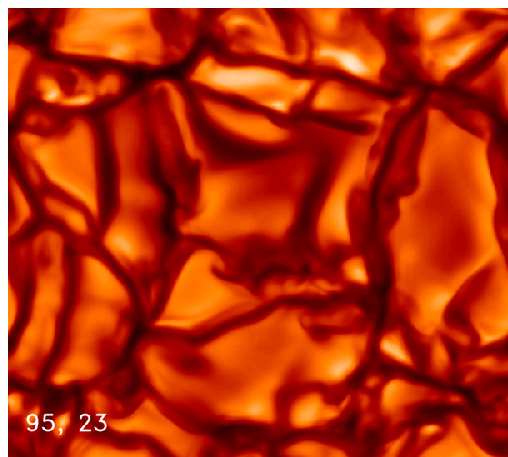}~\includegraphics[angle=90,width=45mm]{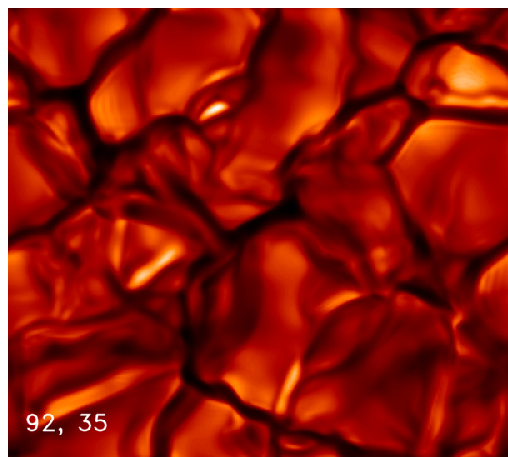}~\includegraphics[angle=90,width=45mm]{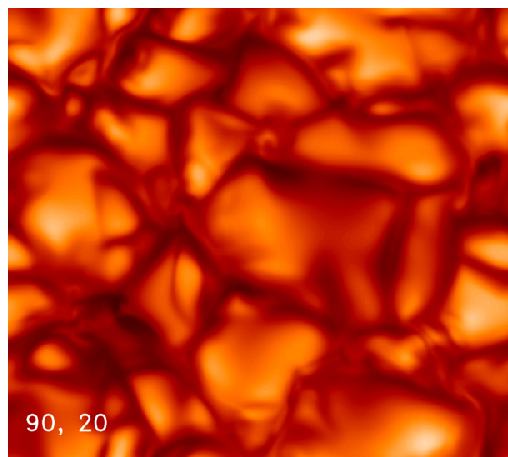}

\vspace{0.001mm}

\includegraphics[angle=90,width=45mm]{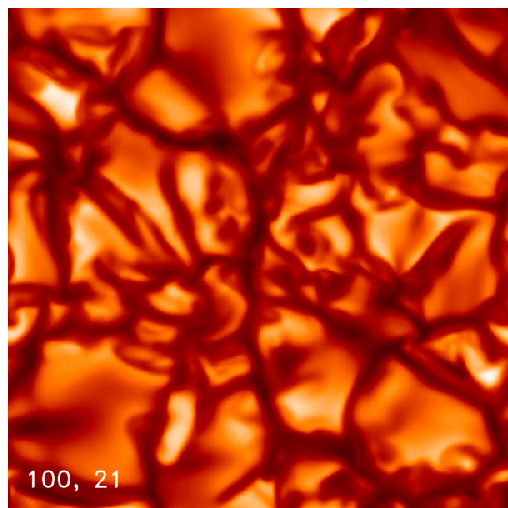}~\includegraphics[angle=90,width=45mm]{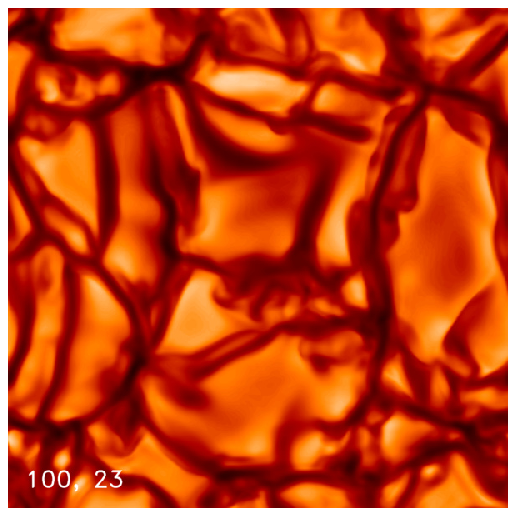}~\includegraphics[angle=90,width=45mm]{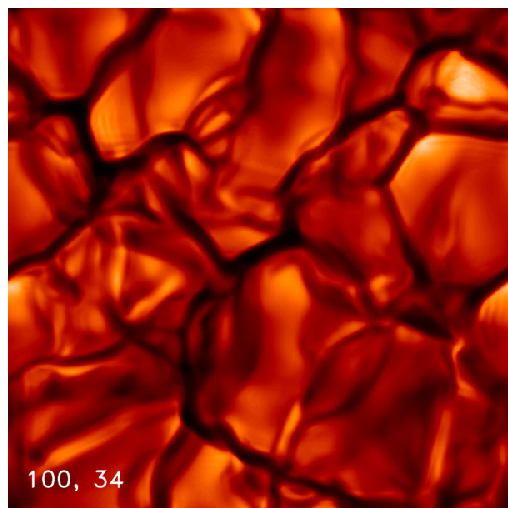}~\includegraphics[angle=90,width=45mm]{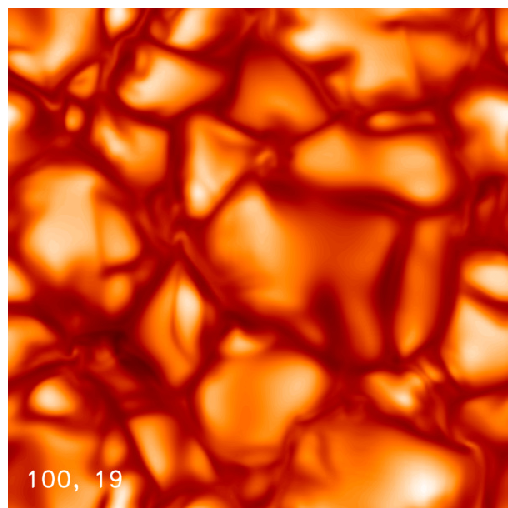}

\hspace{17mm}$8.0\,\left[\mathrm{Mm}\right]$\hspace{30mm}$28.0\,\left[\mathrm{Mm}\right]$\hspace{30mm}$2400.0\,\left[\mathrm{Mm}\right]$\hspace{29mm}$1.4\,\left[\mathrm{Mm}\right]$

\caption{The emergent monochromatic intensity at $500\,\mathrm{nm}$ shown
for eight $\mu$ angles: 0.2, 0.3, 0.5, 0.7, 0.8, 0.9, 1.0 (\emph{from
top to down, respectively}) and for a selection of stars: main-sequence,
turnoff, K-giant, K-dwarf (\emph{from left to right, respectively})
with solar metallicity. Furthermore, we indicated the normalized mean
intensity and the intensity contrast (both in percent).}
\label{fig:intensity_maps}
\end{figure*}
\begin{figure*}
\includegraphics[width=176mm]{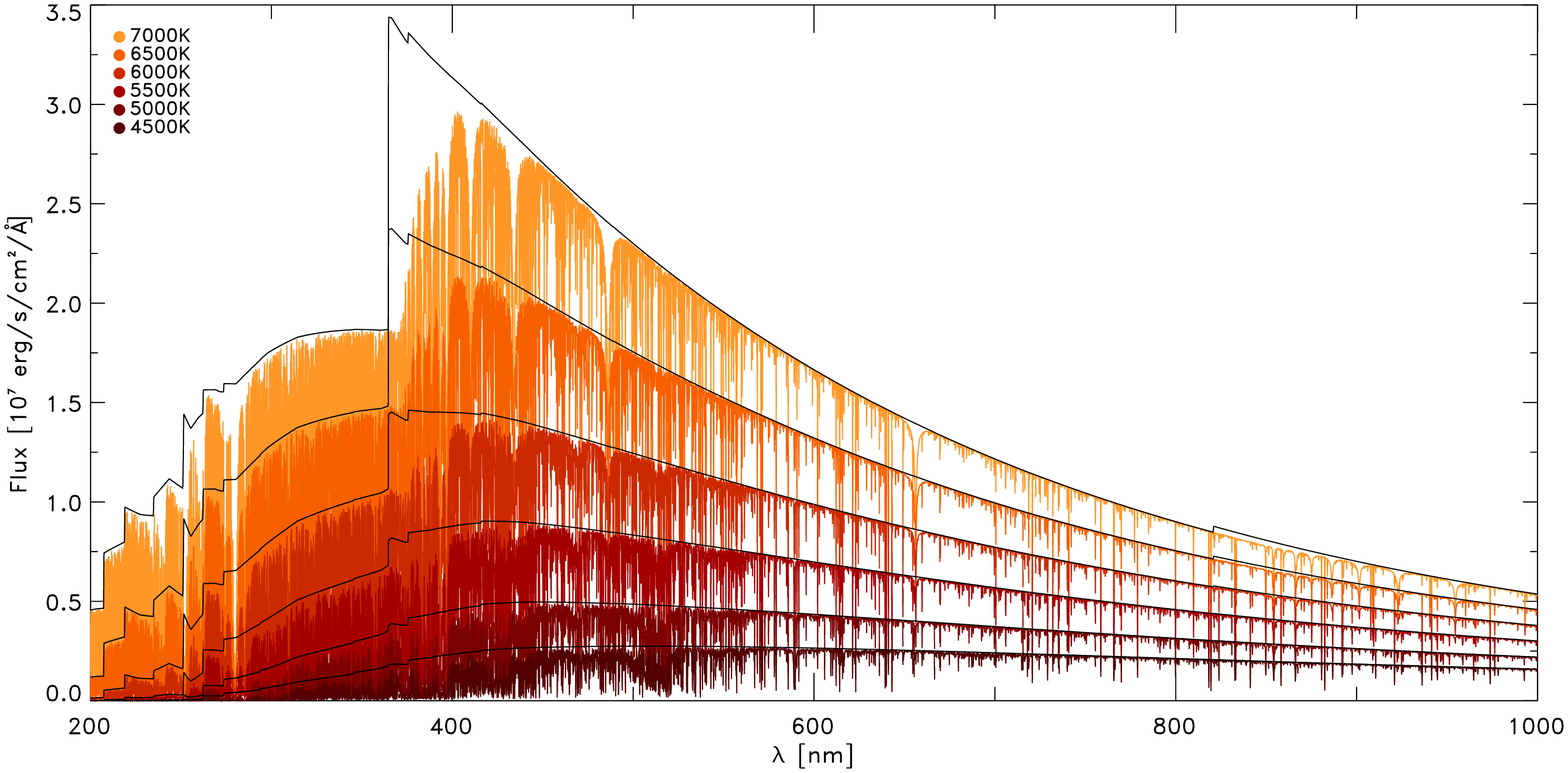}

\caption{Spectral energy distribution vs. wavelength for models with $\logg=4.5$,
$\feh=0.0$ and various $\teff$ (red/orange colors). The continuum
fluxes are also shown (solid black lines).}

\label{fig:flux}
\end{figure*}
Based on the 3D RHD models from the \textsc{Stagger}-grid, we computed
a comprehensive library of full 3D synthetic spectra. Therefore, we
used the \textsc{Optim3D}-code \citep[see][for further details]{Chiavassa:2009p22491,Chiavassa:2010p6257},
which is a post-processing 3D radiative transfer code that assumes
local thermodynamic equilibrium ($S_{\lambda}\left(T\right)=B_{\lambda}\left(T\right)$).
The code considers the realistic velocity field due to convective
motions present in the 3D RHD simulations, thereby taking Doppler
broadening and shifts into account. \textsc{Optim3D} employs pre-tabulated
extinction coefficients are the same as in the MARCS and \textsc{Stagger}
codes, thereby accounting for continuous and sampled line opacities
\citep{Gustafsson:2008p3814}. We assumed the latest solar composition
by \citet{Asplund:2009p3308} consistently in \textsc{Optim3D} as
in the 3D RHD simulations performed with the \textsc{Stagger}-code.
In contrast to the RHD-code, the spectral synthesis code computes
the large number of wavelength points with $N_{\lambda}=105\,767$
explicitly, thereby raising the computational costs enormously. We
achieve a wavelength resolution with a constant sampling rate of $\lambda/\Delta\lambda=20\,000$,
however, we cover a broad range with $\lambda=1010.0$ to $199\,960.0\,\mathring{A}$.
We apply $N_{t}=10$ snapshots for each simulation, while we keep
the horizontal mesh resolution fixed with $N_{xy}=240$. Four (equidistant)
azimutal $\phi$-angles are considered ($N_{\phi}=4$), while the
center-to-limb resolution is resolved higher with nine $\mu$-angles
besides the disk-center ($\mu=0.01,0.05,0.10,0.20,0.30,0.50,0.70,0.80,0.90,1.00$).
After carrying out tests with more $\mu$-angles, we found the ten
chosen angles being sufficient to resolve the limb darkening accurately
at lower computational costs. The strongest decline in the limb darkening
is usually found towards the limb, therefore, we decided to resolve
the limb more instead of equidistant $\mu$-angles. As shown by \citet{Hayek:2012p21944}
the numerical resolution of our 3D RHD models and the resolution for
the spectral flux computations are sufficient to predict realistic
observed limb darkening laws accurately. The synthetic spectra will
be discussed in a separate work (Chiavassa et al. in perp.). 

We show an overview of spatially resolved intensity maps with different
inclined $\mu$ angles for a selection of distinct stellar parameters
(see Fig. \ref{fig:intensity_maps}). These exhibit the typical granulation
pattern of cool stars due to convection. The bright, bulk regions
are the hotter upflowing granules, which are interspersed with the
dark intergranular downdrafts. From the disk-center towards the limb,
the brightness is diminishing significantly, and the intensity contrast
is also slightly dropping. Moreover, one can also obtain that bright
features are often highly angle dependent.

In Fig. \ref{fig:flux}, we show a subset of the resulting averaged
synthetic fluxes in the range $2000-10\,000\,\mathring{A}$ for a
number of dwarfs with solar metallicity. Furthermore, we show also
the continuum fluxes as well, and one can discern spectral absorption
features, the prominent one being the Balmer lines (indicated in the
figure). For higher $\teff$ the continuum flux is increasing, while
individual spectral absorption features are changing as well.

\section{Deriving the limb darkening\label{sec:Deriving-limb-darkening}}

The variation of the inclination of the line of sight from the disk-center
is parameterized with the projected polar angle, $\mu=\cos\theta$,
where $\theta$ is the angle between the line of sight and the direction
of the emergent radiation. Therefore, the disk-center is depicted
with $\mu=1$, while $\mu=0$ is the limb. The limb darkening law
is expressed as the variation in intensity with $\mu$-angle that
is normalized to the disk-center, i.e. $I\left(\mu\right)/I\left(1\right)$.
The resulting monochromatic intensity depends on the horizontal position
$x$ and $y$, the viewing angles $\phi$ and $\mu$ and the time,
$t$, thus $I_{\lambda}\left(x,y,\mu,\phi,t\right)$. In order to
yield the mean monochromatic intensity $\left\langle I_{\lambda}\right\rangle \left(\mu\right)$
from the latter, we average the intensity first spatially, then over
the azimutal angles, and finally over all time steps, i.e. 

\begin{eqnarray*}
\left\langle I_{\lambda}\right\rangle \left(x,y,\mu,\phi,t\right) & = & \frac{1}{N_{t}}\sum_{t}\frac{1}{N_{\phi}}\sum_{\phi}\frac{1}{N_{x}N_{y}}\sum_{x,y}I_{\lambda}\left(x,y,\mu,\phi,t\right).
\end{eqnarray*}
As next, we compute the inclination-dependent total emergent intensity
$I\left(\mu\right)$ by integrating the mean monochromatic intensity
$\left\langle I_{\lambda}\right\rangle $ over all wavelength points
with
\begin{eqnarray*}
I\left(\mu\right) & = & \int\left\langle I_{\lambda}\right\rangle \left(\mu\right)\, d\lambda.
\end{eqnarray*}
Then, the total surface brightness variation can be easily derived
by normalizing the angular intensities with the disk-center value,
$I_{\mu}/I_{1}$, and we can fit the various (bi-parametric) functional
bases, 
\begin{eqnarray}
I_{\mu}/I_{1} & = & 1-u\left(1-\mu\right),\label{eq:linear}\\
I_{\mu}/I_{1} & = & 1-a\left(1-\mu\right)-b\left(1-\mu\right)^{2},\label{eq:quadratic}\\
I_{\mu}/I_{1} & = & 1-c\left(1-\mu\right)-d\left(1-\sqrt{\mu}\right),\label{eq:square_root}\\
I_{\mu}/I_{1} & = & 1-e\left(1-\mu\right)-f\mu\ln\mu,\label{eq:logarithmic}\\
I_{\mu}/I_{1} & = & 1-g\left(1-\mu\right)-h/\left(1-e^{\mu}\right),\label{eq:exponential}
\end{eqnarray}
which are the linear (Eq. \ref{eq:linear}), quadratic (Eq. \ref{eq:quadratic}),
square root (Eq. \ref{eq:square_root}), logarithmic (Eq. \ref{eq:logarithmic})
and exponential (Eq. \ref{eq:exponential}) limb darkening law. Also
the three-parameter non-linear limb darkening law, 
\begin{eqnarray}
I_{\mu}/I_{1} & = & 1-a_{2}(1-\mu)-a_{3}(1-\mu^{3/2})-a_{4}(1-\mu^{2}),\label{eq:three_param}
\end{eqnarray}
introduced by \citet{Sing:2010p21961} is also considered. However,
we recommend the use of the standard four-parameter non-linear functional
basis (Eq. \ref{eq:four-param}) introduced by \citet{Claret:2000p12465},
which is the default limb darkening law in the present study. The
four-parameter power law is the fourth order Taylor-series expansion
in $\mu^{1/2}$ given by

\begin{eqnarray}
I_{\mu}/I_{1} & = & 1-\sum_{k=1}^{4}a_{k}\left(1-\mu^{k/2}\right),\label{eq:four-param}
\end{eqnarray}
This functional basis conserves the flux to better than $0.05\,\%$
\citep[see][]{Claret:2000p12465}. In order to fit the LDC, we applied
the Levenberg-Marquardt least-square minimization, since \citet{Claret:2000p12465}
showed that this fitting method performs best.

\begin{figure}
\includegraphics[width=88mm]{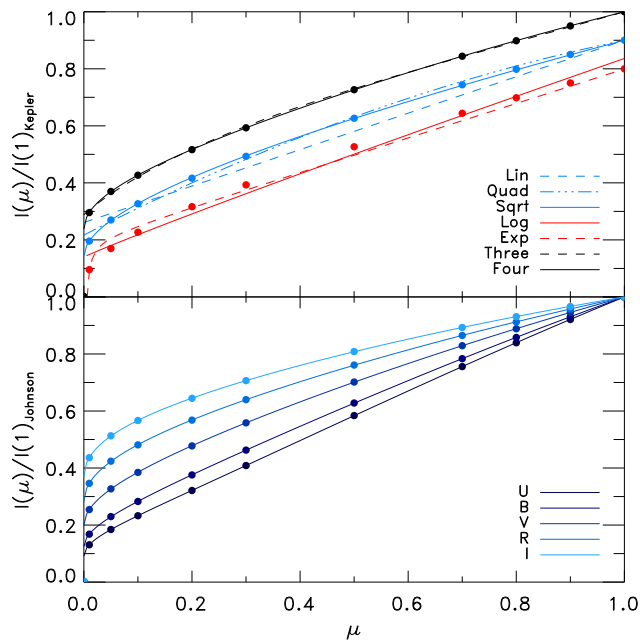}

\caption{Solar intensity distribution vs. $\mu$-angle for different limb darkening
laws in the Kepler filter (top) and different bands from the Johnson
filters (bottom panel). The 3D results are indicated with filled circles.
In the top panel the blue and red lines are shifted by $0.1$ and
$0.2$ respectively, while the black lines are unshifted.}

\label{fig:sun_ld}
\end{figure}
\begin{table}
\caption{\label{tab:precision}The precision of the functional fits for the
different limb darkening laws (see text for details).}

\begin{tabular}{ccccccc} 
\hline\hline
Value & Lin & Quad & Sqrt & Log & Three & Four \\
\hline
          $\overline{\chi^2}$  &       1.53 &       1.20 &       0.72 &      13.38 &       2.13 &       2.35$\times 10^{-6}$ \\
     $\overline{\max \delta}$  &       1.20 &       1.79 &       1.56 &       2.00 &       1.37 &       1.65$\times 10^{-3}$ \\
$\overline{\mu}$ &       0.07 &       0.05 &       0.05 &       0.08 &       0.06 &       0.10 \\
\hline\end{tabular}
\end{table}
To illustrate the performance of the individual functional basis we
show in Fig. \ref{fig:sun_ld} (top panel) the limb darkening laws
of the solar simulation seen in the Kepler filter. The two-coefficient
laws are obviously rather inadequate and show the largest deviations
at the limb ($\mu\sim0.0$), in particular the linear, quadratic,
logarithmic and exponential law (Eqs. \ref{eq:linear}, \ref{eq:quadratic},
\ref{eq:logarithmic} and \ref{eq:exponential}), while the square
root law (Eq. \ref{eq:square_root}) exhibits a rather good match
(it is already known that the square root law performs better for
hotter stars, while for cooler stars the quadratic law is better see
\citealt{Claret:2000p12465}). The three-parameter functional basis
(Eq. \ref{eq:three_param}) is performing well, however, it mismatches
the limb rather significantly. The standard four-parameter non-linear
power law (Eq. \ref{eq:four-param}) is an excellent functional basis,
and due to its versatility the fits result in extreme small residuals.
In order to depict the precision of the individual laws quantitatively,
we list the average $\chi^{2}$, average maximal relative deviation
and average location of the latter from all stellar models in Table
\ref{tab:precision}. The four-parameter law performs for all stellar
parameters significantly much better than any other limb darkening
law, therefore, we will discuss subsequently the latter only.

We consider a number of broad band filters $b$ by convolving the
response function $S_{\lambda}$, which considers the transmission
of the filter $b$, with the integration of the intensity, 
\begin{eqnarray*}
I_{b}\left(\mu\right) & = & \int S_{\lambda}\left\langle I_{\lambda}\right\rangle \left(\mu\right)\, d\lambda.
\end{eqnarray*}
We applied multiple standard broad band filters taken from the SYNPHOT
package%
\footnote{\url{http://www.stsci.edu}%
}, which comprises Bessel (JHK), Johnson (UBVRI) and Str\"omgren (uvby).
Additionally we considered individual important instruments with CoRoT,
Kepler%
\footnote{\url{http://keplergo.arc.nasa.gov/}%
}, Mauna Kea (JHKLM), SDSS (ugriz) and HST (ACS, STIS). Furthermore,
the complete grid of spectra will be made available online, such that
limb darkening from different filters or wavelength bands can be derived.
In Fig. \ref{fig:sun_ld} (bottom panel) we show the five different
Johnson filters for the solar model. The brightness distribution and
its curvature are becoming more enhanced towards higher wavelength
from the ultra-violet to the infra-red, which is a general feature
for all stellar parameters. The optical depth and the temperature
gradient are dependent on the considered wavelength, since radiation
at higher wavelength is emerging from higher geometrical depth.

\section{Limb darkening\label{sec:Limb-darkening-laws}}

\begin{figure}
\includegraphics[width=88mm]{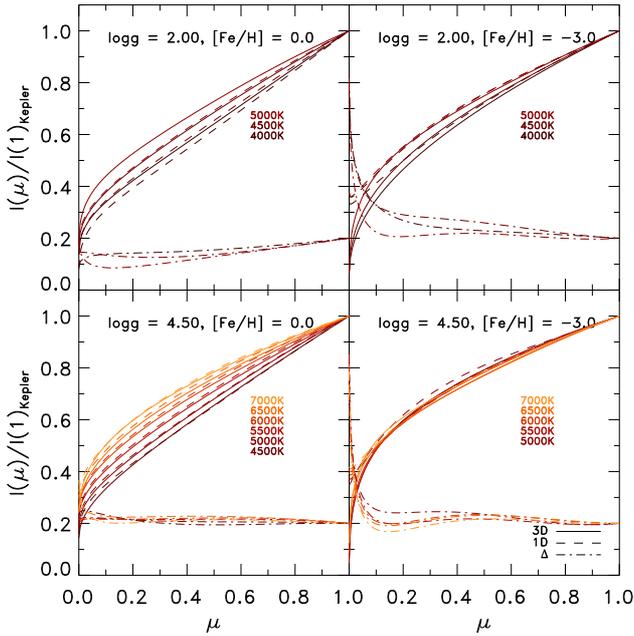}

\caption{Disk-center normalized intensity, $I\left(\mu\right)/I\left(1\right)$,
against the inclination, $\mu$, in the Kepler filter showing the
limb darkening for different stellar parameters. Furthermore, we illustrate
also the 1D ATLAS predictions for comparison (dashed lines), and the
deviations $\Delta=1\mathrm{D}-3\mathrm{D}$ (dashed dotted lines),
which are enhanced by a factor of $2$ and shifted by $+0.2$ for
clarity.}
\label{fig:ov_limb-dark}
\end{figure}
\begin{figure*}
\includegraphics[width=35.2mm]{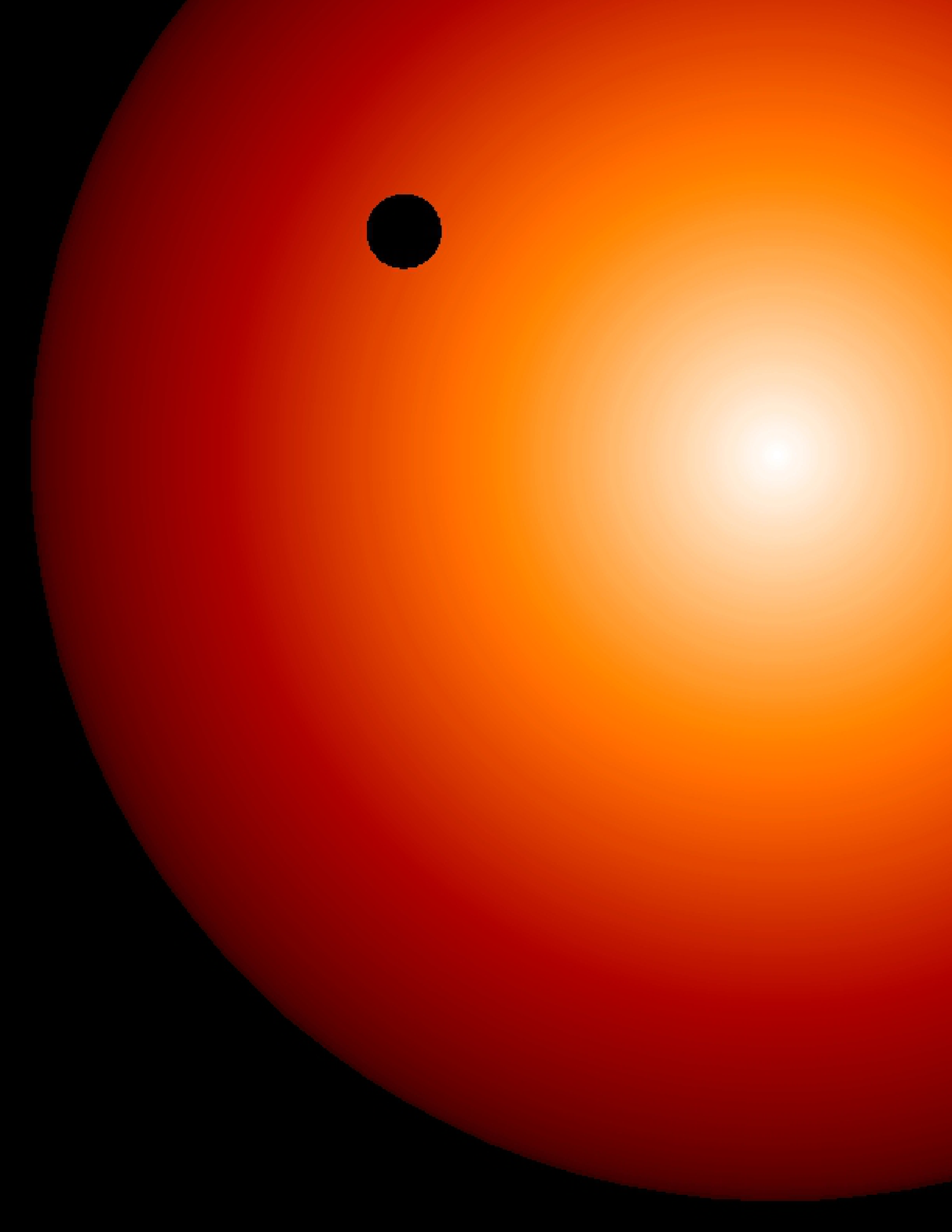}~~\includegraphics[width=35.2mm]{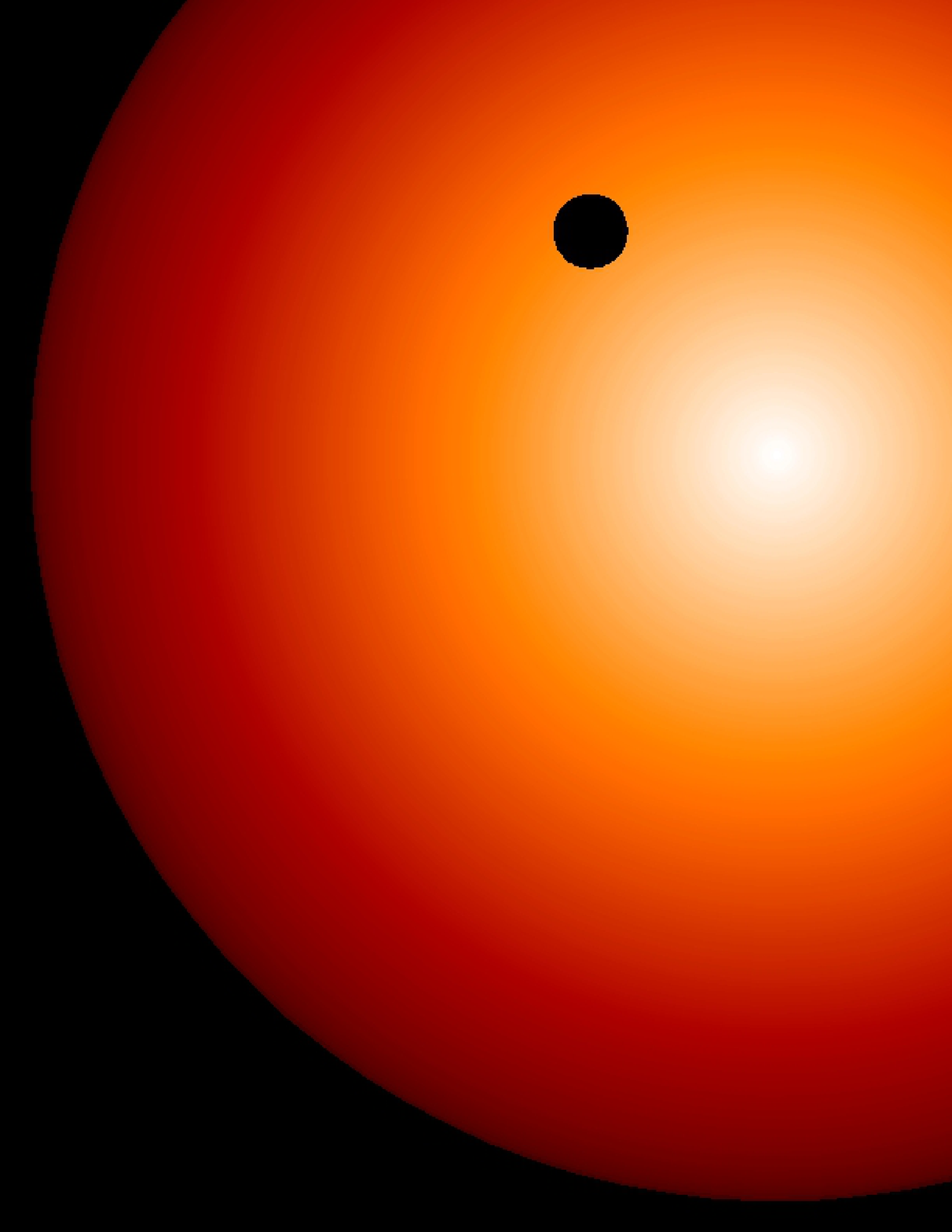}~~\includegraphics[width=35.2mm]{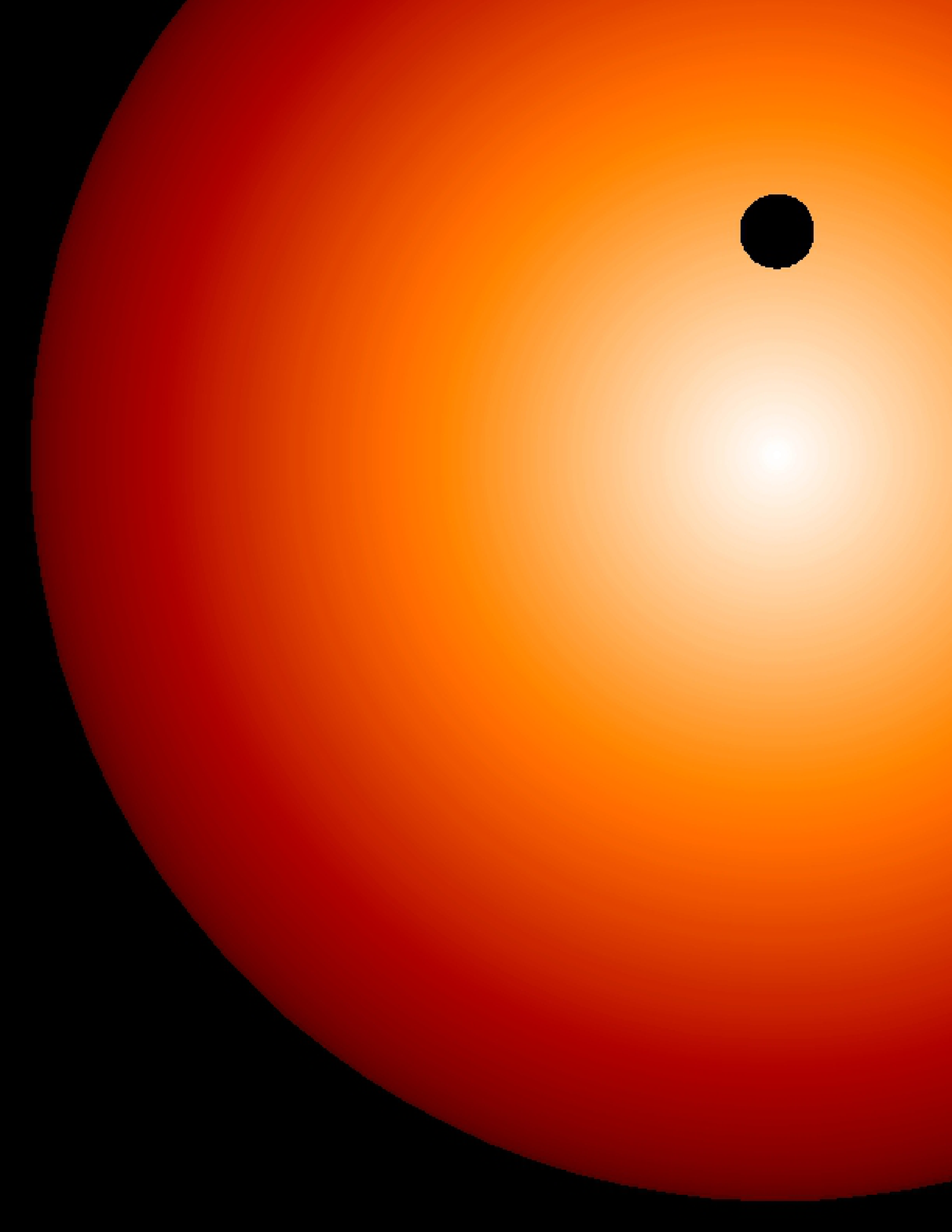}~~\includegraphics[width=35.2mm]{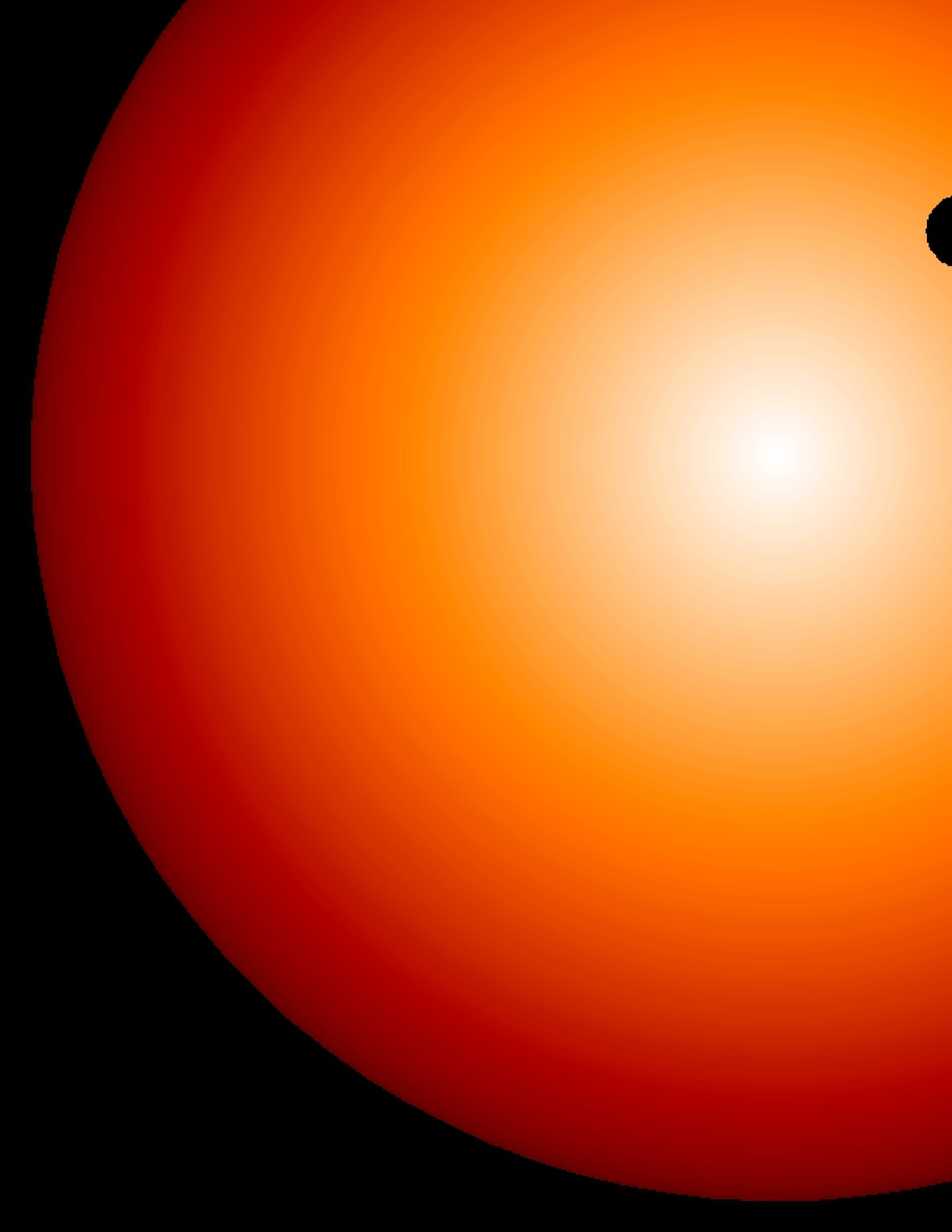}~~\includegraphics[width=35.2mm]{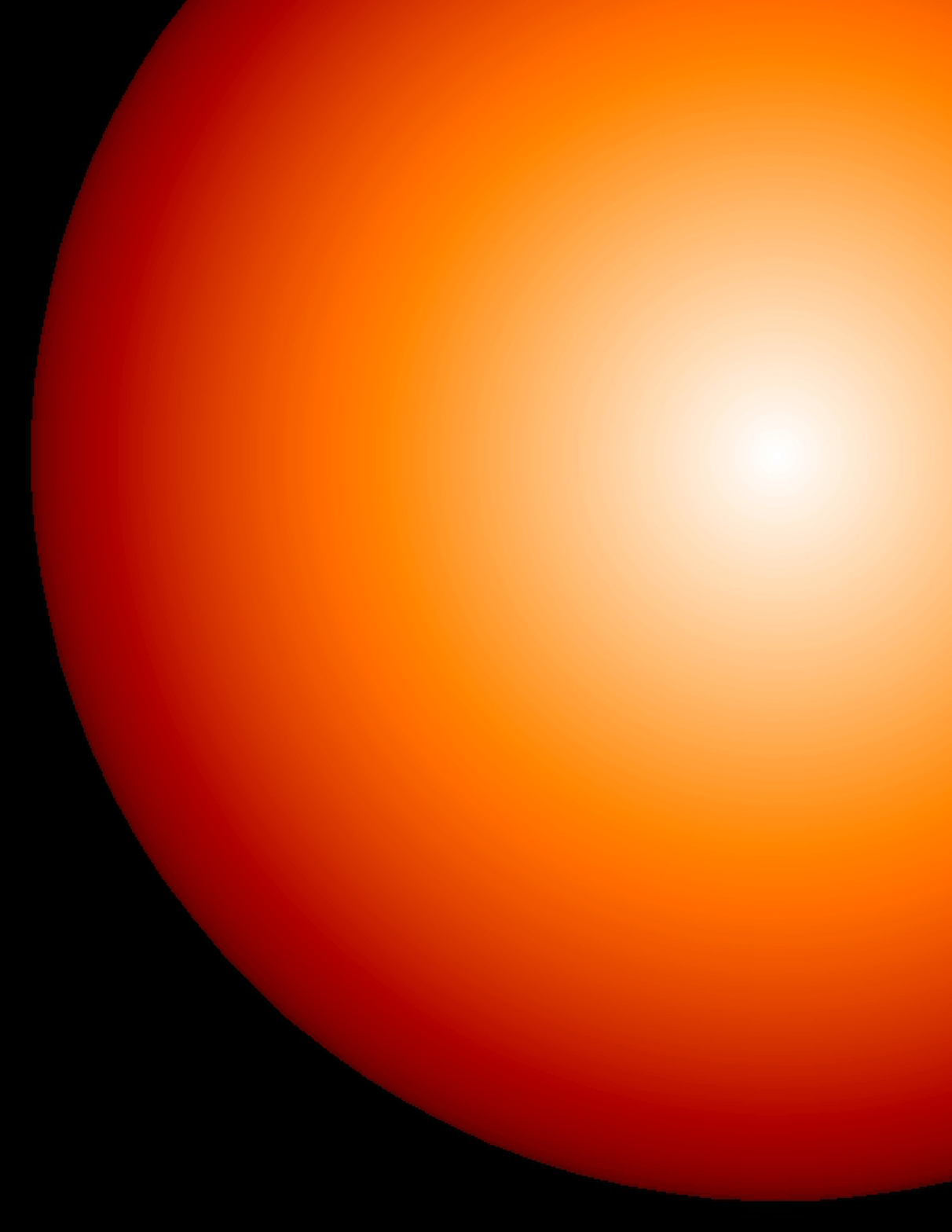}

\caption{Synthetic stellar disks for stars with $\logg=4.5$, $\feh=0$, and
increasing $\teff$ from $4500\,\mathrm{K}$ to $6500\,\mathrm{K}$
(from left to right) seen in the Kepler filter, and including a transiting
exoplanet with $p=0.05$ with progressing transit phase. Note the
brighter limb towards higher $\teff$.}
\label{fig:synthetic_disk}
\end{figure*}
The emergent radiative intensity of stars decreases almost linear
monotonically from the center to the edge, until it drops-off sharply
close to the limb ($\mu\sim0.1$), which is known as limb darkening
and can be observed in the Sun. The radiation at disk-center emerges
from lower depths, while towards the limb one observes light from
higher layers, where the temperature stratification has dropped very
quickly, hence lower (darker) brightness ($I\propto T^{4}$). We mention
that the limb darkening has obvious boundary constraints being that
the intensity is maximal with $I\left(\mu=1\right)=1$ and minimal
with $I\left(\mu=0\right)\simeq0$. Furthermore, we note briefly that
the variation of the limb darkening with stellar parameter depends
also on the considered filter, and the results can differ significantly
(Fig. \ref{fig:sun_ld}).

In Fig. \ref{fig:ov_limb-dark}, we show an overview of different
limb darkening variations for various stellar parameters. One finds
in general that the variations are rather smooth and systematic, and
the largest differences are given close to the limb ($\mu\sim0.2$),
however, the variations with $\teff$ and $\logg$ are distinctive
for different metallicities. For higher $\teff$ the decline in brightness
exhibits a more pronounced convex curvature between $\mu=0.2-0.7$,
while at cooler ones it is close to a linear drop. Therefore, hotter
models end up with brighter intensities towards the limb around $\mu\sim0.2$
and a steeper drop beyond. For instance for dwarfs with solar-metallicity
we find $I_{0.2}/I_{1}=0.4-0.6$ for $\teff=4500-7000\,\mathrm{K}$.
In order to illustrate the relative brightening at the limb for hotter
dwarfs with solar metallicity, we show synthetic stellar disks in
Fig. \ref{fig:synthetic_disk} with higher $\teff$. Towards giants
(lower $\logg$) the brightness is higher than for dwarfs, however
the changes are more subtle compared to the effective temperature.
At lower metallicity, we find the differences with $\teff$ between
metal-poor dwarfs ($\logg=4.5,$ $\feh=-3$) being distinctively smaller,
so that a pronounced curvature is given even for the coolest effective
temperature (compare bottom panels in Fig. \ref{fig:ov_limb-dark}).
In fact, we find basically no increase with $\teff$ around $\mu\sim0.2$
($I_{0.2}/I_{1}=0.59-0.61$). As we show further down, the $T$-insensitivity
of the limb darkening at $\feh=-3$ arises due to the temperature
gradient. \citet{Claret:2000p12465} had also found an enhanced curvature
at lower metallicity with 1D models. The hottest and most metal-poor
dwarfs are the brightest at the edge, and the sharp drop is the steepest.
Furthermore, we note that the center-to-limb variation curves for
metal-poor models cross the corresponding ones for solar-metallicity
models with otherwise the same stellar parameters ($\teff$), with
the exception of the models at high $\teff$, however, this is not
the case towards higher $\teff$. On the other hand, the metal-poor
giants are more similar to the solar-metallicity case, and exhibit
also a clear $\teff$ sensitivity.

\begin{figure}
\includegraphics[width=88mm]{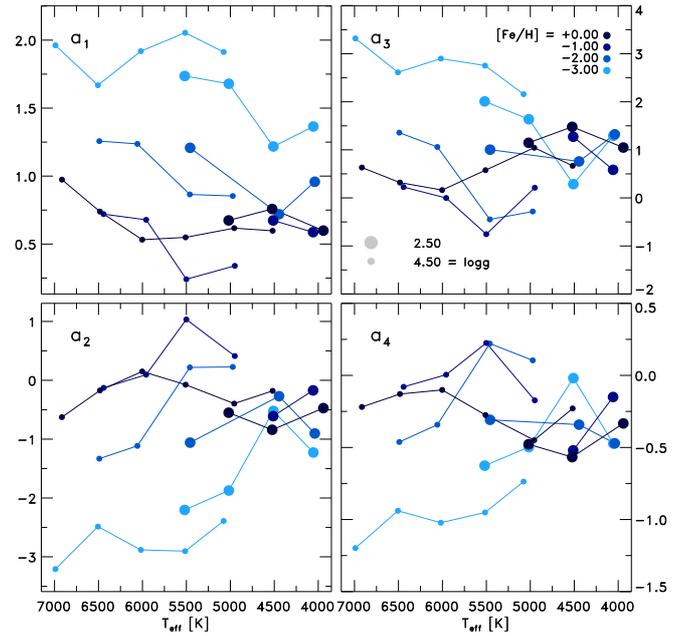}

\caption{The four coefficients for the four-parameter non-linear limb darkening
law (Eq. \ref{eq:four-param}) predicted in the Kepler filter from
3D RHD models for different stellar parameters.}
\label{fig:four_coefficients}
\end{figure}
As next, we want to discuss the individual coefficients for the four-parameter
limb darkening law (Eq. \ref{eq:four-param}). Therefore, we show
in Fig. \ref{fig:four_coefficients} the four coefficients for different
$\teff$ and $\feh$ and $\logg=4.5,2.0$. Towards higher $\teff$
both coefficients, $a_{2}$ and $a_{4}$, are increasing until $6000\,\mathrm{K}$,
then above they decrease, while the coefficients $a_{1}$ and $a_{3}$
vary the opposite at solar metallicity. For lower $\feh$ the $T$-dependence
is inverted for $a_{k}$ and the even coefficients, $a_{2}$ and $a_{4}$,
are decreasing, and the odd ones, $a_{1}$ and $a_{3}$, are increasing.
Another aspect worthy of attention is the correlation between the
coefficients with half-integer exponents in $\mu$ ($a_{1}$ and $a_{3}$),
and integer one ($a_{2}$ and $a_{4}$) with $\teff$ and $\feh$
(compare left with right panels in Fig. \ref{fig:four_coefficients}).
In the Kepler filter we find the correlations to amount with $C\left[a_{1},a_{3}\right]=0.82$
and $C\left[a_{2},a_{4}\right]=0.91$ for all stellar parameters.
Furthermore, the half-integer  exponents anti-correlate with the integer
ones (compare top with bottom panels in Fig. \ref{fig:four_coefficients}).
While the distant coefficients are less anti-correlated with $C\left[a_{1},a_{4}\right]=-0.76$,
we find the anti-correlation between the successive coefficients being
much tighter with $C\left[a_{1},a_{2}\right]=-0.95$ and $C\left[a_{3},a_{4}\right]=-0.99$.
\citet{Claret:2000p12465} noted also correlations between the coefficients.
The coefficients $a_{k}$ of the four-parameter law can be decomposed
and considered individually. Then the integer exponents with the coefficients
$a_{2}$ and $a_{4}$ are leading to a linear and quadratic polynomial
respectively, which are describing the general slope of the limb darkening.
On the other hand, the half-integer exponents with the coefficients
$a_{1}$ and $a_{3}$ are square root like functions, and are responsible
for the curvature towards the limb.

\begin{figure}
\includegraphics[width=88mm]{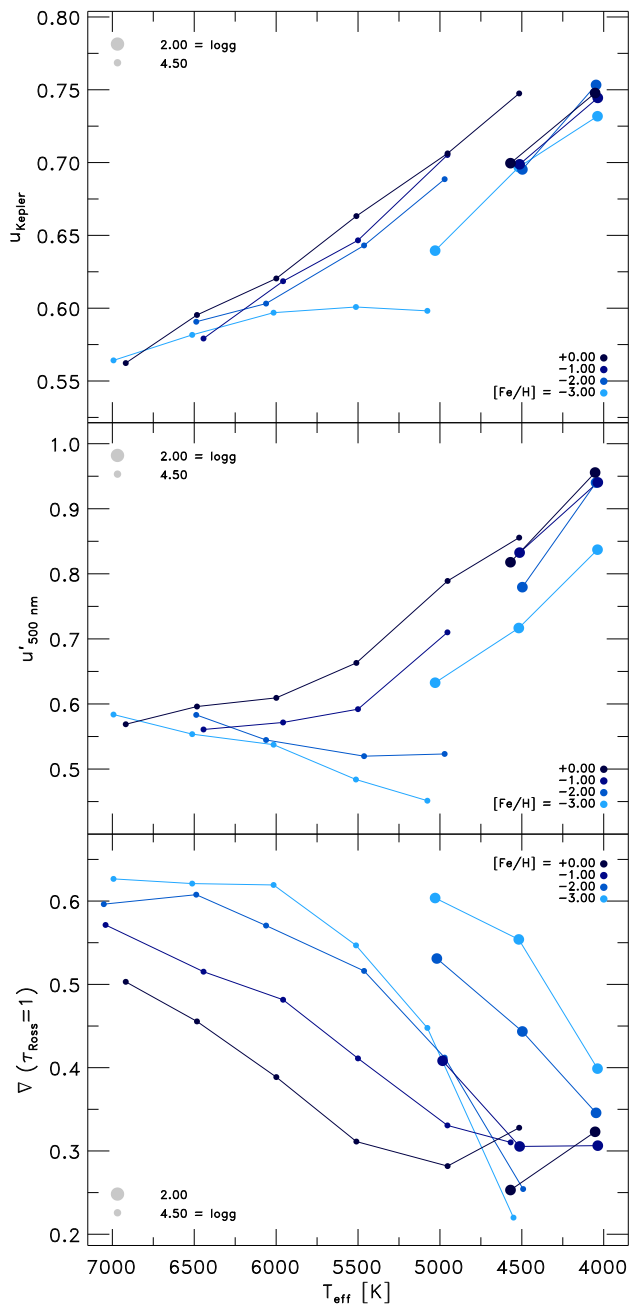}

\caption{The coefficient $u$ of the linear limb darkening law (Eq. \ref{eq:linear})
in the Kepler filter (top) and the approximation $u_{500\,\mathrm{nm}}^{'}$
from Eq. \ref{eq:lin_u_approx} (middle) and temperature gradient
at the optical surface (bottom panel) vs. $\teff$ for $\logg=2.0,\,4.5$
and different $\feh$.}
\label{fig:lin-u_nab}
\end{figure}
The coefficient $u$ for the linear limb darkening law (Eq. \ref{eq:linear})
is rather crude, however, it has the major advantage of simplicity,
since the linear law reduces the complex shape of the limb darkening
into a single value. We display in Fig. \ref{fig:lin-u_nab} $u$
against $\teff$ for different stellar parameters (top panel). A larger
value in $u$ relates to a steeper drop in intensity and indicate
lower brightness at the limb, and a lower value for $u$ results in
brighter limbs (see Eq. \ref{eq:linear}). The coefficient $u$ is
mainly sensitive to the effective temperature and is decreasing with
higher $\teff$, and for different $\logg$ and $\feh$ the differences
are rather small. As a remark we note that the linear coefficients
$a$ and $c$ from Eqs. \ref{eq:quadratic} and \ref{eq:square_root}
are similar, while the quadratic and square root coefficients, $b$
and $d$, behave oppositely and increase for higher $\teff$ (not
shown). The values for $u$ ranges from 0.56 to 0.77 in the Kepler
filter, which would yield $I=0.44$ and $0.23$ at the limb ($\mu=0$)
respectively (the global range in $u$ is from $0.03$ to 1.05). 

With a linear approximation of the Planck function one can derive
the slope of the linear limb darkening law, $u$ \citep[e.g.][]{Gray2005oasp.book.....G,Hayek:2012p21944},
which depends primarily on the temperature gradient at the optical
surface and is given by 
\begin{eqnarray}
u_{\lambda}^{'} & \approx & \frac{\log e}{B_{\lambda}\left(T\left(\tau_{\lambda}=1\right)\right)}\frac{dB_{\lambda}}{dT}\Big|_{\tau_{\lambda}=1}\frac{dT}{d\log\tau}\Big|_{\tau_{\lambda}=1}\label{eq:lin_u_approx}
\end{eqnarray}
The approximation implies that a steeper temperature gradient at the
optical surface will lead to stronger (steeper) limb darkening (larger
$u$). In Fig. \ref{fig:lin-u_nab} (middle panel) we show also $u_{500\,\mathrm{nm}}^{'}$
considered at the optical surface ($\tau=1$) vs. $\teff$, and we
find $u_{500\,\mathrm{nm}}^{'}$ to correlate well with $u$ (top
panel). Also, similar as given in $u$, we find $u_{500\,\mathrm{nm}}^{'}$
being $T$-insensitive for dwarf models with very low metallicity
($\feh=-3$), which arises from the temperature gradient term $dT/d\log\tau_{\lambda}$
in Eq. \ref{eq:lin_u_approx}. The intensity is given by the source
function and the lost radiation, which is stated by the radiative
transfer eq., $dI_{\lambda}/d\tau_{\lambda}=S_{\lambda}-I_{\lambda}$.
Under the assumption of LTE, the source function can be approximated
with the Planck function at the local temperature, $S_{\lambda}\left(T\right)=B_{\lambda}\left(T\right)$.
Therefore, the variation of the intensity with $\mu$ is sensitive
to the temperature structure and in particular the temperature gradient.
In Fig. \ref{fig:lin-u_nab} we show also the temperature gradient,
$\nab=d\ln T/d\ln p_{\mathrm{tot}}$, considered at the optical surface
(bottom panel). For lower metallicity the range in temperature gradient
is enhanced, which is very similar to the intensity contrast. We had
already mentioned the enhancement of the intensity contrast and temperature
gradient at lower metallicity in Paper I. We find the reason for the
enhancement being the lack of metals that are usually the most important
electron donors for the formation of $\mathrm{H}^{-}$, which is the
dominating opacity source \citep[see][]{Nordlund:1990p6720}. Therefore,
in metal-poor models, the main contribution of electrons arises from
the ionization of hydrogen. This is the reason for the strong enhancement
of the of temperature gradient towards $\teff=6000\,\mathrm{K}$ with
$\feh=-3$, which is the reason for the $T$-insensitivity of $u$
and the limb darkening that we found for metal-poor dwarfs (see Fig.
\ref{fig:ov_limb-dark}).

\section{Transit light curves\label{sec:Transit-light-curves}}

\begin{figure*}
\subfloat[\label{fig:light_curve}]{\includegraphics[width=88mm]{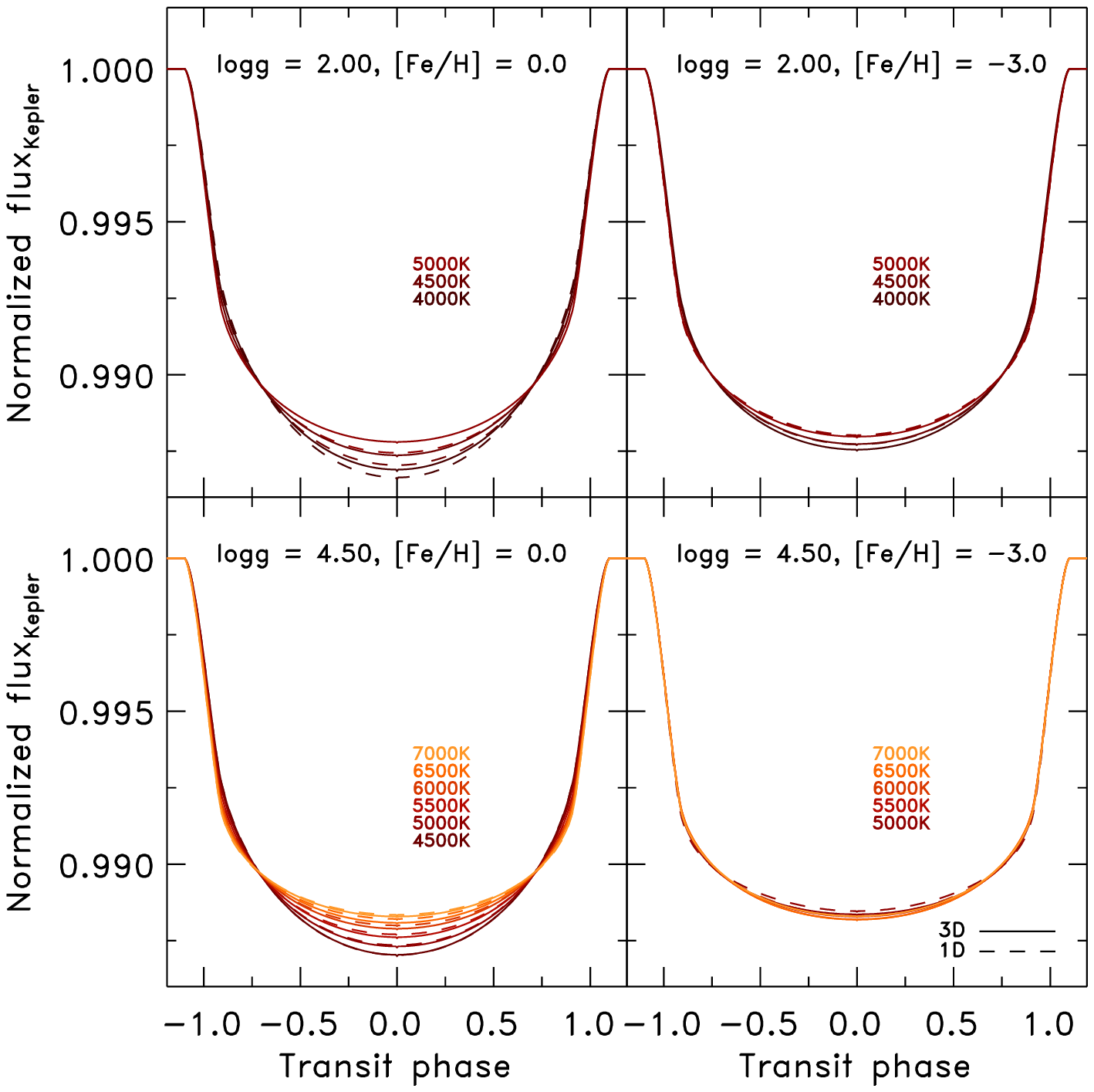}

}\subfloat[\label{fig:light_curve_diff1d}]{\includegraphics[width=88mm]{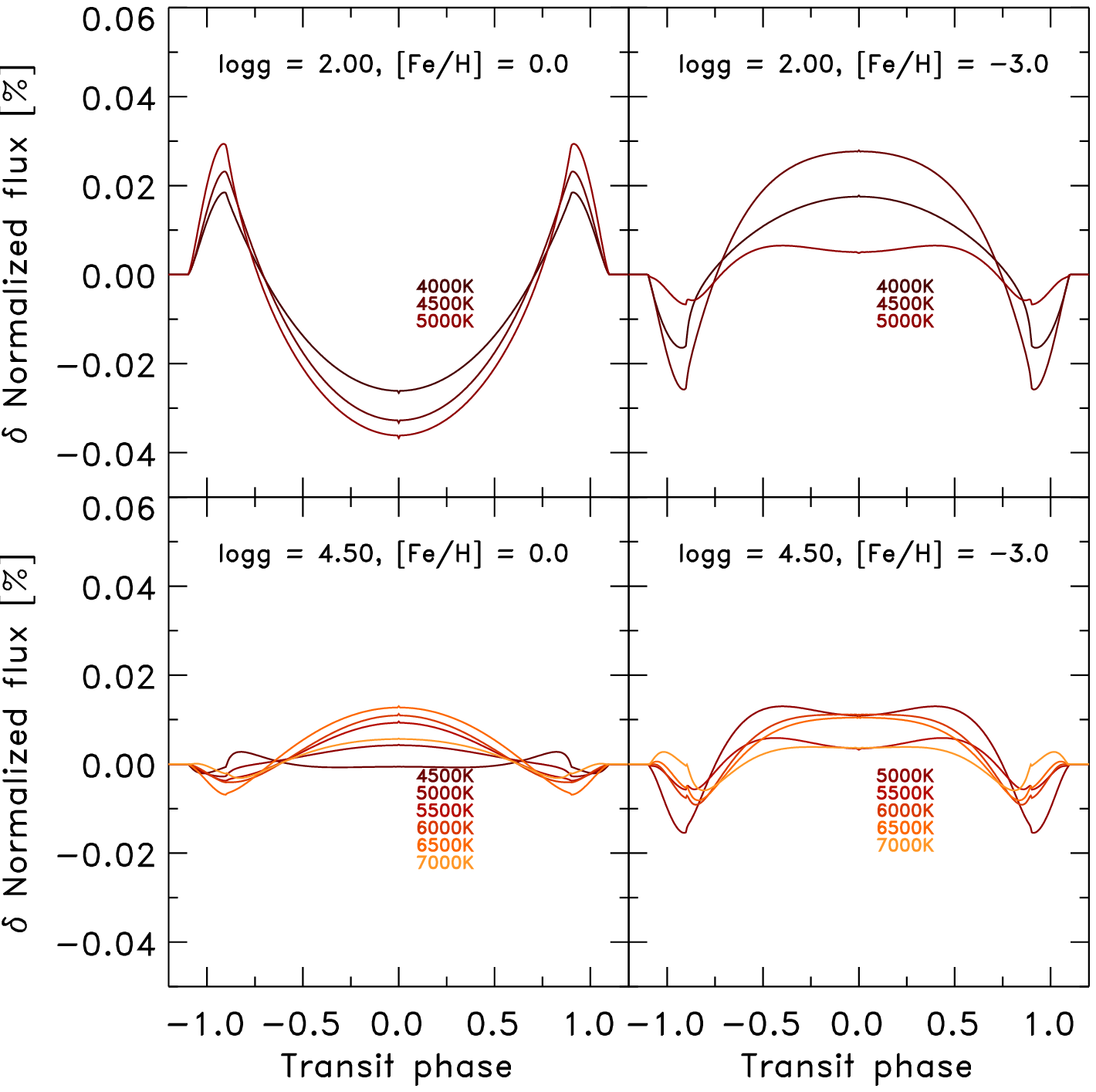}

}

\caption{\emph{Left figure}: transit light curve vs. transit phase with $p=0.1$
in the Kepler filter for different stellar parameters. The predictions
from the 1D ATLAS models are also included (dashed lines). \emph{Right
figure}: relative deviations in the transit light curve with $p=0.1$
between 3D atmosphere models and 1D ATLAS models given in $\%$. The
difference is $\delta=1\mathrm{D}/3\mathrm{D}-1$.}
\end{figure*}
During exo-solar planet transits the planet eclipses its host star
in the line of sight to earth, thereby diminishing the emergent intensity
and leaving a characteristic imprint in the observed light curve.
Theoretical light curve predictions are described by two main parameters,
namely the ratio of the radii of planet and parent star, $p=r_{\mathrm{planet}}/r_{\mathrm{star}}$,
and the normalized separation of the centers, $z=d/r_{\mathrm{star}}$,
with $d$ being the center-to-center distance of the two occulting
bodies. For computing the theoretical transit light curves, we used
the publicly available code by \citet{Mandel:2002p22255}. In Fig.
\ref{fig:light_curve}, we show light curves with $p=0.1$, i.e. for
the case that the radius of the planet is $1/10$ of the star, in
the Kepler filter for the same LDC as discussed in Sec. \ref{sec:Limb-darkening-laws}.
Then one can extract that the dwarfs with solar-metallicity exhibit
for higher $\teff$ light curves with a more rectangular shaped feature,
where the transit center is shallower and the width of the transit
is broader than lower $\teff$. The giant models show similar features
with $\teff$ for both $\feh$. On the other hand, the metal-poor
dwarfs are similarly box-shaped and do not alter much for lower effective
temperatures. The limb of the intensity distribution will shape the
ingress and egress of the light curve, e.g. when the limb darkening
would be a step function ($a_{2}\simeq0.$ and $a_{k}=0.$), then
the light curve would be entirely rectangular. A straight linear dropping
intensity distribution ($a_{2}=1$ and $a_{k}=0.$) results in a more
elliptical shape with a narrow width, while a curved square root drop
($a_{1}=1$ and $a_{k}=0.$) would lead to an evenly circular shaped
light curve. Therefore, we caution for the use of limb darkening laws
that are incapable of rendering the drop-off at the limb, such as
the bi- and three-parametric laws, since these will introduce inevitably
systematic errors in the theoretical transit light curves. \citet{Mandel:2002p22255}
found in a comparison between the quadratic and the four-parameter
non-linear power law (with $p=0.1$) differences by $3\,\%$! Furthermore,
the depth of the transit light curve depends primarily on the ratio
of planet to host $p$. For larger $p$ the light curves are increasingly
deeper at the center, since more stellar light is effectively blocked
during the transition due to a larger surface ratio of the planet.
The limb darkening is also sensitive to the considered wavelength-regime
(see Fig. \ref{fig:sun_ld}). Therefore, the transit light curve will
be different depending on the actual considered broad band filter,
which samples its specific range in $\lambda$. We find in general
that the light curve is towards lower wavelength (ultra-violet) systematically
more convex shaped with a deeper center and more slender width, while
towards higher wavelength (infra-red) the light curve is more box-shaped
with shallower centers and broader width\@. We note that a multi-band
photometry approach states a solution to this issue \citep[see][]{Knutson:2007p22139}.

\section{Comparison with results from 1D models\label{sec:Comparison-with-1D}}

\begin{figure}
\includegraphics[width=88mm]{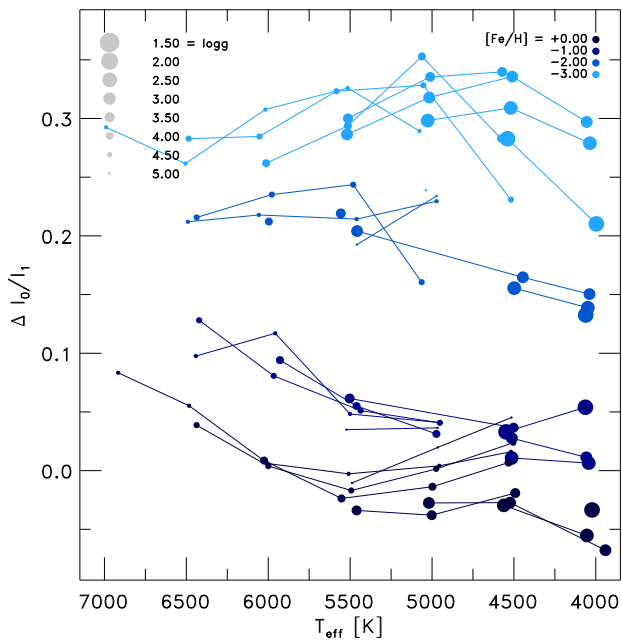}

\caption{Overview of the maximal relative difference in the limb darkening
compared between 1D and 3D against $\teff$ with the Kepler filter.
The relative deviations are retrieved by $\Delta=1\mathrm{D}-3\mathrm{D}$}

\label{fig:ov_lddiff}
\end{figure}
The ATLAS models are the widest applied 1D atmosphere models for retrieving
LDC, since its grid covers the broadest range in stellar parameters
leading to seamless coverage. The differences between the 1D and 3D
models arise mainly from the differences in the temperature stratification,
in particular, the temperature gradients. 

For the comparison, we show in Fig. \ref{fig:ov_limb-dark} also the
limb darkening derived from 1D MLT models from the ATLAS grid \citep[see][]{Sing:2010p21961}.
To ensure consistency the four-parameter non-linear laws for 1D models
are also shown in the Kepler filter. The 1D dwarf models with solar
metallicity exhibit similar increasing curvatures for higher $\teff$
as given by the 3D models, only the 1D models are slightly brighter
than the 3D, except for the coolest one ($4500\,\mathrm{K}$). In
the case of giants the 1D limb darkening is distributed at much lower
and also more linear intensities. The limb of the metal-poor 1D models
lacks of a similar smooth sharp drop-off that is given in the 3D models,
instead they depict a rather discontinuous behavior at the limb, which
is doubtful to be correct. For metal-poor models it is known that
the enforcement of radiative equilibrium is leading to an overestimation
of the temperature stratification in the upper layers due to lack
of spectral line absorption \citep[see][]{Asplund:1999p11771,Collet:2007p5617}.
The radiation at the limb is emerging from higher layers, therefore,
the deviations at the limb are consistent (see also Paper I).

The largest deviations are usually found at the limb, since some of
the 1D models depict an almost linear run at the edge, while the 3D
models exhibit a comparably smooth drop-off at the limb. In Figure
\ref{fig:ov_lddiff}, we show the deviations at the limb, $\Delta I_{0}/I_{1}$,
between the 1D and 3D limb darkening predictions. The differences
increase significantly towards lower metallicity. Therefore, we advise
the use of 3D limb darkening coefficients in particular for metal-poor
stars.

We display the relative differences in the transit light curve between
the 3D and 1D predictions in Fig. \ref{fig:light_curve_diff1d}. Similar
to the findings of \citet{Hayek:2012p21944}, we find a characteristic
shape in the residuals with two extrema that are usually of opposite
sign. One being at the ingress and egress phase, which arises from
differences at the limb, and the other one taking place at the disk-center.
We found in Sec. \ref{sec:Limb-darkening-laws} the 1D models being
brighter than the 3D limb darkening, except for the giants at solar-metallicity.
The transit light curves for the 1D models are similarly brighter
at the disk-center and dimmer at the transit phase edges for all models,
only for the solar-metallicity giants it is the opposite. However,
the maximal residuals are relatively small with a global maximum of
$\sim0.04\,\%$ in the transit light curve.

Since the differences in the transit light curves between 3D and 1D
are rather small, the expected differences in transit parameters are
also small. \citet{Hayek:2012p21944} determined a larger (smaller)
radius of the exoplanet HD 209458b (HD 189733b) by 0.2 \% in the wavelength
region between 2900 and 5700 $\mathring{A}$. Hence, a reanalysis
of the observed transit light curves with our LDC might change the
transit parameters less than a percent.

\section{Conclusions\label{sec:Conclusions}}

We derived on the basis of the \textsc{Stagger}-grid, a large grid
of 3D RHD atmosphere models, the limb darkening coefficients for the
various bi-parametric and non-linear limb darkening laws. The four-parameter
non-linear power law introduced by \citet{Claret:2000p12465} is the
only limb darkening law that is sufficiently versatile to express
the intensity distribution with an excellent accuracy, while all other
limb darkening laws are insufficient, in particular at the limb. Therefore,
we recommend the use of the four-parameter functional basis only,
in particular for the comparison with high-precision measurements
in the hunt of extrasolar planets. We discussed the limb darkening
in the Kepler filter for various stellar parameters, and outlined
systematical variations that exposed the complex changes of the brightness
distribution, in particular with the effective temperature. We compared
also our new LDC with predictions from widely used 1D ATLAS models,
and the largest differences are given towards the limb. The 1D models
are often brighter than 3D predictions, only for giant models with
solar-metallicity we find opposite differences. Furthermore, we displayed
the systematic (anti-)correlations between the coefficients $a_{k}$
between half-integer and integer exponents of the four-parameter law.
We found the coefficient of linear limb darkening law, $u$, to scale
with the temperature gradient and the Planck function. Theoretical
transit light curves indicate similar systematical differences between
1D and 3D as the limb darkening variations implied, which are relatively
small. However, as observations indicate \citep{Knutson:2007p22139,Hayek:2012p21944},
these can be measured with high-precision observations. Therefore,
we advise to use of of the new LDC.
\begin{acknowledgements}
We acknowledge access to computing facilities at the Rechenzentrum
Garching (RZG) of the Max Planck Society and at the Australian National
Computational Infrastructure (NCI) where the simulations were carried
out. Also, we acknowledge the Action de Recherche Concert{\'e}e (ARC)
grant provided by the Direction g{\'e}n{\'e}rale de l’Enseignement
non obligatoire et de la Recherche scientifique – Direction de la
Recherche scientifique – Communaut{\'e} française de Belgique, and
the F.R.S.- FNRS. Remo Collet is the recipient of an Australian Research
Council Discovery Early Career Researcher Award (project number DE120102940).
\end{acknowledgements}
\bibliographystyle{aa}
\bibliography{papers}

\begin{thebibliography}{41}
\expandafter\ifx\csname natexlab\endcsname\relax\def\natexlab#1{#1}\fi

\bibitem[{{Allende Prieto} {et~al.}(2002){Allende Prieto}, {Asplund},
  {Garc{\'{\i}}a L{\'o}pez}, \& {Lambert}}]{AllendePrieto:2002p22280}
{Allende Prieto}, C., {Asplund}, M., {Garc{\'{\i}}a L{\'o}pez}, R.~J., \&
  {Lambert}, D.~L. 2002, \apj, 567, 544

\bibitem[{{Asplund} {et~al.}(2009){Asplund}, {Grevesse}, {Sauval}, \&
  {Scott}}]{Asplund:2009p3308}
{Asplund}, M., {Grevesse}, N., {Sauval}, A.~J., \& {Scott}, P. 2009, \araa, 47,
  481

\bibitem[{{Asplund} {et~al.}(1999){Asplund}, {Nordlund}, {Trampedach}, \&
  {Stein}}]{Asplund:1999p11771}
{Asplund}, M., {Nordlund}, {\AA}., {Trampedach}, R., \& {Stein}, R.~F. 1999,
  \aap, 346, L17

\bibitem[{{Aufdenberg} {et~al.}(2005){Aufdenberg}, {Ludwig}, \&
  {Kervella}}]{Aufdenberg2005ApJ...633..424A}
{Aufdenberg}, J.~P., {Ludwig}, H.-G., \& {Kervella}, P. 2005, \apj, 633, 424

\bibitem[{{Bigot} {et~al.}(2006){Bigot}, {Kervella}, {Th{\'e}venin}, \&
  {S{\'e}gransan}}]{Bigot:2006p23513}
{Bigot}, L., {Kervella}, P., {Th{\'e}venin}, F., \& {S{\'e}gransan}, D. 2006,
  \aap, 446, 635

\bibitem[{{B{\"o}hm-Vitense}(1958)}]{BohmVitense:1958p4822}
{B{\"o}hm-Vitense}, E. 1958, \zap, 46, 108

\bibitem[{{Charbonneau} {et~al.}(2000){Charbonneau}, {Brown}, {Latham}, \&
  {Mayor}}]{Charbonneau:2000p22468}
{Charbonneau}, D., {Brown}, T.~M., {Latham}, D.~W., \& {Mayor}, M. 2000, \apjl,
  529, L45

\bibitem[{{Chiavassa} {et~al.}(2012){Chiavassa}, {Bigot}, {Kervella}, {Matter},
  {Lopez}, {Collet}, {Magic}, \& {Asplund}}]{Chiavassa:2012p22493}
{Chiavassa}, A., {Bigot}, L., {Kervella}, P., {et~al.} 2012, \aap, 540, A5

\bibitem[{{Chiavassa} {et~al.}(2010){Chiavassa}, {Collet}, {Casagrande}, \&
  {Asplund}}]{Chiavassa:2010p6257}
{Chiavassa}, A., {Collet}, R., {Casagrande}, L., \& {Asplund}, M. 2010, \aap,
  524, A93

\bibitem[{{Chiavassa} {et~al.}(2009){Chiavassa}, {Plez}, {Josselin}, \&
  {Freytag}}]{Chiavassa:2009p22491}
{Chiavassa}, A., {Plez}, B., {Josselin}, E., \& {Freytag}, B. 2009, \aap, 506,
  1351

\bibitem[{{Claret}(2000)}]{Claret:2000p12465}
{Claret}, A. 2000, \aap, 363, 1081

\bibitem[{{Claret}(2004)}]{Claret:2004p12494}
{Claret}, A. 2004, \aap, 428, 1001

\bibitem[{{Claret} {et~al.}(1995){Claret}, {Diaz-Cordoves}, \&
  {Gimenez}}]{Claret:1995p22394}
{Claret}, A., {Diaz-Cordoves}, J., \& {Gimenez}, A. 1995, \aaps, 114, 247

\bibitem[{{Collet} {et~al.}(2007){Collet}, {Asplund}, \&
  {Trampedach}}]{Collet:2007p5617}
{Collet}, R., {Asplund}, M., \& {Trampedach}, R. 2007, \aap, 469, 687

\bibitem[{{Davis} {et~al.}(2000){Davis}, {Tango}, \&
  {Booth}}]{Davis2000MNRAS.318..387D}
{Davis}, J., {Tango}, W.~J., \& {Booth}, A.~J. 2000, \mnras, 318, 387

\bibitem[{{Diaz-Cordoves} {et~al.}(1995){Diaz-Cordoves}, {Claret}, \&
  {Gimenez}}]{DiazCordoves:1995p22374}
{Diaz-Cordoves}, J., {Claret}, A., \& {Gimenez}, A. 1995, \aaps, 110, 329

\bibitem[{{Gray}(2005)}]{Gray2005oasp.book.....G}
{Gray}, D.~F. 2005, {The Observation and Analysis of Stellar Photospheres}

\bibitem[{{Gustafsson} {et~al.}(2008){Gustafsson}, {Edvardsson}, {Eriksson},
  {J{\o}rgensen}, {Nordlund}, \& {Plez}}]{Gustafsson:2008p3814}
{Gustafsson}, B., {Edvardsson}, B., {Eriksson}, K., {et~al.} 2008, \aap, 486,
  951

\bibitem[{{Hayek} {et~al.}(2012){Hayek}, {Sing}, {Pont}, \&
  {Asplund}}]{Hayek:2012p21944}
{Hayek}, W., {Sing}, D., {Pont}, F., \& {Asplund}, M. 2012, \aap, 539, A102

\bibitem[{{Knutson} {et~al.}(2007){Knutson}, {Charbonneau}, {Noyes}, {Brown},
  \& {Gilliland}}]{Knutson:2007p22139}
{Knutson}, H.~A., {Charbonneau}, D., {Noyes}, R.~W., {Brown}, T.~M., \&
  {Gilliland}, R.~L. 2007, \apj, 655, 564

\bibitem[{{Kurucz}(1979)}]{Kurucz:1979p4707}
{Kurucz}, R.~L. 1979, \apjs, 40, 1

\bibitem[{{Magic} \& {Asplund}(2014)}]{Magic2014arXiv1405.7628M}
{Magic}, Z. \& {Asplund}, M. 2014, ArXiv e-prints

\bibitem[{{Magic} {et~al.}(2014{\natexlab{a}}){Magic}, {Collet}, \&
  {Asplund}}]{Magic2014arXiv1403.6245M}
{Magic}, Z., {Collet}, R., \& {Asplund}, M. 2014{\natexlab{a}}, ArXiv e-prints

\bibitem[{{Magic} {et~al.}(2013{\natexlab{a}}){Magic}, {Collet}, {Asplund},
  {Trampedach}, {Hayek}, {Chiavassa}, {Stein}, \& {Nordlund}}]{Magic:2013}
{Magic}, Z., {Collet}, R., {Asplund}, M., {et~al.} 2013{\natexlab{a}}, \aap,
  557, A26

\bibitem[{{Magic} {et~al.}(2013{\natexlab{b}}){Magic}, {Collet}, {Hayek}, \&
  {Asplund}}]{magic:2013arXiv1307.3273M}
{Magic}, Z., {Collet}, R., {Hayek}, W., \& {Asplund}, M. 2013{\natexlab{b}},
  \aap, 560, A8

\bibitem[{{Magic} {et~al.}(2014{\natexlab{b}}){Magic}, {Weiss}, \&
  {Asplund}}]{Magic2014arXiv1403.1062M}
{Magic}, Z., {Weiss}, A., \& {Asplund}, M. 2014{\natexlab{b}}, ArXiv e-prints

\bibitem[{{Mandel} \& {Agol}(2002)}]{Mandel:2002p22255}
{Mandel}, K. \& {Agol}, E. 2002, \apjl, 580, L171

\bibitem[{{Mayor} \& {Queloz}(1995)}]{Mayor:1995p23118}
{Mayor}, M. \& {Queloz}, D. 1995, \nat, 378, 355

\bibitem[{{Mihalas}(1970)}]{Mihalas:1970p21310}
{Mihalas}, D. 1970

\bibitem[{{Milne}(1921)}]{Milne1921MNRAS..81..361M}
{Milne}, E.~A. 1921, \mnras, 81, 361

\bibitem[{{Nordlund}(1982)}]{Nordlund:1982p6697}
{Nordlund}, A. 1982, \aap, 107, 1

\bibitem[{{Nordlund} \& {Dravins}(1990)}]{Nordlund:1990p6720}
{Nordlund}, A. \& {Dravins}, D. 1990, \aap, 228, 155

\bibitem[{{Nordlund} {et~al.}(2009){Nordlund}, {Stein}, \&
  {Asplund}}]{Nordlund:2009p4109}
{Nordlund}, {\AA}., {Stein}, R.~F., \& {Asplund}, M. 2009, Living Reviews in
  Solar Physics, 6, 2

\bibitem[{{Pereira} {et~al.}(2013){Pereira}, {Asplund}, {Collet}, {Thaler},
  {Trampedach}, \& {Leenaarts}}]{Pereira:2013arXiv1304}
{Pereira}, T.~M.~D., {Asplund}, M., {Collet}, R., {et~al.} 2013, \aap, 554,
  A118

\bibitem[{{Seager} \& {Sasselov}(2000)}]{Seager:2000p14757}
{Seager}, S. \& {Sasselov}, D.~D. 2000, \apj, 537, 916

\bibitem[{{Sing}(2010)}]{Sing:2010p21961}
{Sing}, D.~K. 2010, \aap, 510, A21

\bibitem[{{Skartlien}(2000)}]{Skartlien:2000p9857}
{Skartlien}, R. 2000, \apj, 536, 465

\bibitem[{{Southworth}(2008)}]{Southworth:2008p23114}
{Southworth}, J. 2008, \mnras, 386, 1644

\bibitem[{{Stein} \& {Nordlund}(1998)}]{Stein:1998p3801}
{Stein}, R.~F. \& {Nordlund}, A. 1998, \apj, 499, 914

\bibitem[{{van Hamme}(1993)}]{vanHamme:1993p22403}
{van Hamme}, W. 1993, \aj, 106, 2096

\bibitem[{{Wright} {et~al.}(2011){Wright}, {Fakhouri}, {Marcy}, {Han}, {Feng},
  {Johnson}, {Howard}, {Fischer}, {Valenti}, {Anderson}, \&
  {Piskunov}}]{Wright2011PASP..123..412W}
{Wright}, J.~T., {Fakhouri}, O., {Marcy}, G.~W., {et~al.} 2011, \pasp, 123, 412

\end{thebibliography}
\appendix

\section{Table with the limb darkening coefficients\label{sec:Table-LDC}}

In Table \ref{tab:limb_dark_coeff} we listed a subset of the limb
darkening coefficients for F, G, and K main-sequence stars. The full
table is available at CDS.
\begin{table*}
\caption{\label{tab:limb_dark_coeff} Limb darkening coefficients derived from
3D RHD models in the Kepler filter for different stellar parameters
(Cols. 1,2,3). The linear (Col. 4), quadratic (Cols. 5,6), square
root (Cols. 7,8) and four-parameter non-linear laws (Cols. 9,10,11,12)
are listed.}
\begin{tabular}{lllllllllllll}
\hline\hline
$T_{\rm{eff}}$ & $\log g$ & [Fe/H] & $u$ & $a$ & $b$ & $c$ & $d$ & $a_1$ & $a_2$ & $a_3$ & $a_4$\\
\hline
      4532 &    3.50 &    +0.0 &    0.73505 &    0.58775 &    0.17369 &    0.47757 &    0.33638 &    0.77661 &   -0.80075 &    1.38617 &   -0.50608\\
      4997 &    3.50 &    +0.0 &    0.69317 &    0.51184 &    0.21380 &    0.40246 &    0.37978 &    0.63495 &   -0.46625 &    1.10377 &   -0.46910\\
      5552 &    3.50 &    +0.0 &    0.63205 &    0.37638 &    0.30146 &    0.23583 &    0.51764 &    0.52385 &    0.09984 &    0.29682 &   -0.16943\\
      6025 &    3.50 &    +0.0 &    0.57909 &    0.27389 &    0.35988 &    0.08275 &    0.64845 &    0.70830 &    0.00368 &   -0.03424 &    0.06155\\
      4504 &    4.00 &    +0.0 &    0.74641 &    0.58971 &    0.18481 &    0.47456 &    0.35520 &    0.76201 &   -0.69272 &    1.24804 &   -0.44840\\
      4974 &    4.00 &    +0.0 &    0.69924 &    0.51916 &    0.21233 &    0.40603 &    0.38305 &    0.69226 &   -0.60023 &    1.22834 &   -0.50452\\
      5493 &    4.00 &    +0.0 &    0.64892 &    0.42136 &    0.26832 &    0.29549 &    0.46173 &    0.54134 &   -0.07649 &    0.58178 &   -0.28535\\
      5999 &    4.00 &    +0.0 &    0.61169 &    0.34095 &    0.31926 &    0.18721 &    0.55457 &    0.53991 &    0.17288 &    0.08752 &   -0.06136\\
      6437 &    4.00 &    +0.0 &    0.57175 &    0.26414 &    0.36271 &    0.07584 &    0.64788 &    0.80623 &   -0.39503 &    0.52466 &   -0.19731\\
      5767 &    4.44 &    +0.0 &    0.63905 &    0.39256 &    0.29064 &    0.25352 &    0.50367 &    0.58503 &   -0.10226 &    0.53712 &   -0.25796\\
      4518 &    4.50 &    +0.0 &    0.74748 &    0.58106 &    0.19623 &    0.46824 &    0.36481 &    0.59750 &   -0.17959 &    0.66723 &   -0.22919\\
      4955 &    4.50 &    +0.0 &    0.70630 &    0.52186 &    0.21747 &    0.41223 &    0.38419 &    0.61695 &   -0.39597 &    1.04387 &   -0.44957\\
      5509 &    4.50 &    +0.0 &    0.66324 &    0.43841 &    0.26511 &    0.31192 &    0.45897 &    0.54929 &   -0.07493 &    0.57716 &   -0.27516\\
      6002 &    4.50 &    +0.0 &    0.62040 &    0.35566 &    0.31221 &    0.20713 &    0.53998 &    0.53192 &    0.15042 &    0.16292 &   -0.10095\\
      6483 &    4.50 &    +0.0 &    0.59534 &    0.28054 &    0.37119 &    0.09313 &    0.65611 &    0.73898 &   -0.17219 &    0.31894 &   -0.12921\\
      6915 &    4.50 &    +0.0 &    0.56230 &    0.20662 &    0.41939 &   -0.01686 &    0.75662 &    0.97382 &   -0.62622 &    0.63240 &   -0.21903\\
      4515 &    5.00 &    +0.0 &    0.74272 &    0.52349 &    0.25852 &    0.38093 &    0.47270 &    0.66450 &   -0.13643 &    0.51088 &   -0.16603\\
      4965 &    5.00 &    +0.0 &    0.72911 &    0.53095 &    0.23365 &    0.42084 &    0.40274 &    0.47629 &    0.08613 &    0.51634 &   -0.25122\\
      5488 &    5.00 &    +0.0 &    0.66477 &    0.43757 &    0.26789 &    0.31734 &    0.45389 &    0.47390 &    0.09151 &    0.45732 &   -0.25382\\
      5996 &    3.50 &    -2.0 &    0.59147 &    0.23859 &    0.41607 &   -0.00987 &    0.78563 &    1.40907 &   -1.73174 &    1.75496 &   -0.59484\\
      5063 &    4.00 &    -2.0 &    0.67569 &    0.30132 &    0.44143 &    0.09182 &    0.76280 &    0.53808 &    0.75966 &   -0.74362 &    0.27944\\
      5481 &    4.00 &    -2.0 &    0.63979 &    0.22783 &    0.48574 &   -0.03935 &    0.88726 &    1.21877 &   -0.91653 &    0.84285 &   -0.26340\\
      5977 &    4.00 &    -2.0 &    0.59856 &    0.20852 &    0.45991 &   -0.06036 &    0.86084 &    1.43622 &   -1.63025 &    1.57506 &   -0.52309\\
      6437 &    4.00 &    -2.0 &    0.56859 &    0.17828 &    0.46022 &   -0.09412 &    0.86578 &    1.39622 &   -1.46391 &    1.26974 &   -0.37546\\
      5784 &    4.44 &    -2.0 &    0.62103 &    0.20358 &    0.49221 &   -0.06998 &    0.90276 &    1.30219 &   -1.15073 &    1.07186 &   -0.35042\\
      4972 &    4.50 &    -2.0 &    0.68860 &    0.23125 &    0.53926 &   -0.03294 &    0.94265 &    0.85345 &    0.22723 &   -0.28352 &    0.10407\\
      5460 &    4.50 &    -2.0 &    0.64309 &    0.20143 &    0.52075 &   -0.06477 &    0.92477 &    0.86507 &    0.21908 &   -0.44776 &    0.22070\\
      6057 &    4.50 &    -2.0 &    0.60317 &    0.21858 &    0.45348 &   -0.03590 &    0.83491 &    1.23655 &   -1.11440 &    1.05958 &   -0.34213\\
      6490 &    4.50 &    -2.0 &    0.59069 &    0.23563 &    0.41864 &   -0.00428 &    0.77730 &    1.25723 &   -1.33216 &    1.35739 &   -0.46197\\
      4971 &    5.00 &    -2.0 &    0.66775 &    0.17879 &    0.57657 &   -0.10938 &    1.01532 &    1.02230 &   -0.12475 &    0.01055 &   -0.00136\\
      5458 &    5.00 &    -2.0 &    0.64344 &    0.18595 &    0.53943 &   -0.07624 &    0.94022 &    0.66300 &    0.82112 &   -1.09008 &    0.44583\\
\hline
\end{tabular}

\end{table*}

\end{document}